\documentclass[%
 reprint,
superscriptaddress,
nofootinbib,
 amsmath,amssymb,
 aps,
prd,
]{revtex4-2}

\usepackage{xcolor}
\usepackage{graphicx}
\usepackage{dcolumn}
\usepackage{bm}
\usepackage{hyperref}
\usepackage[mathlines]{lineno}
\usepackage[normalem]{ulem}

\usepackage[commentmarkup=uwave]{changes}
\definechangesauthor[name={Will Farr}]{WF}

\usepackage{tikz}

\definecolor{mpl_blue}{HTML}{1F77B4}
\definecolor{mpl_orange}{HTML}{FF7F0E}
\definecolor{mpl_green}{HTML}{2CA02C}
\definecolor{mpl_red}{HTML}{D62728}

\newcommand{\CIT}{\affiliation{Department of Physics, California Institute of Technology, Pasadena, California 91125, USA}}
\newcommand{\CITLab}{\affiliation{LIGO Laboratory, California Institute of Technology, Pasadena, California 91125, USA}}
\newcommand{\CCA}{\affiliation{Center for Computational Astrophysics, Flatiron Institute, 162 5th Ave, New York, New York 10010, USA}}
\newcommand{\StonyBrook}{\affiliation{Department of Physics and Astronomy, Stony Brook University, Stony Brook NY 11794, United States}}

\DeclareMathOperator{\sign}{sign}

\begin{document}

\newcommand{\PopulationAvgOverlapsFromTwoHzMedianRate}{42}
\newcommand{\PopulationAvgOverlapsFromFiveHzMedianRate}{4}
\newcommand{\PopulationAvgOverlapsFromTenHzMedianRate}{0.5}

\newcommand{\PercentOverlapsBelowFiveHz}{91}

\newcommand{\AverageTFNumberOccupiedAtTwoHzMedianRate}{0.91}
\newcommand{\AverageTFNumberOccupiedAtFiveHzMedianRate}{0.17}
\newcommand{\AverageTFNumberOccupiedAtTenHzMedianRate}{0.05}

\newcommand{\TypicalParameterUncertaintyIncrease}{1\%}
\newcommand{\TargetSignalRestrictedUncertainty}{XX}
\newcommand{\DataAfterMergerUncertainty}{XX}

\newcommand{\PercentOfMillionSignals}{0.14\%}

\newcommand{\TrackCrossingsFromTwoHzMedianRate}{41.6}

\newcommand{\MaxTFNumberOccupiedAtTwoHzMedianRate}{6}
\newcommand{\MaxTFNumberOccupiedAtTenHzMedianRate}{3}

\newcommand{\AvgChirpDetector}{$\mathcal{M_z} = 3.93\,M_{\odot}$}
\newcommand{\MedChirpDetector}{$\mathcal{M_z} = 3.6\,M_{\odot}$}
\newcommand{\AvgChirpSource}{$\mathcal{M} = 1.3\,M_{\odot}$}  

\newcommand{\NumOverlapsAvgMedianRateTwoHz}{274.4}

\newcommand{\NumOverlapsAvgChirpSourceZNaught}{$651$}  
\newcommand{\NumOverlapsAvgChirpSourceZOne}{111}

\newcommand{\PopulationAvgOverlapsFromTwoHzLowRate}{9}
\newcommand{\PopulationAvgOverlapsFromTwoHzHighRate}{145}

\newcommand{\DetectorFrameChirpMassVaryPercent}{3}
\newcommand{\MedianDetectorFrameChirpMassVoy}{1.3\,$M_\odot$}
\newcommand{\MedianDetectorFrameChirpMassCE}{2.9\,$M_\odot$}
\newcommand{\MedianDetectorFrameChirpMassET}{2.5\,$M_\odot$}
\newcommand{\MedianDetectorFrameChirpMassCEET}{3.2\,$M_\odot$}

\newcommand{\NumOverlappingOverNumInWindowLowRate}{0.33}
\newcommand{\NumOverlappingOverNumInWindowMedianRate}{0.26}
\newcommand{\NumOverlappingOverNumInWindowHighRate}{0.39}
\newcommand{\AvgNumOverlappingOverNumInWindow}{0.3}

\newcommand{\ExtremeEventRateAvgSignalsPerBinTwoHz}{15}

\newcommand{\SNRSignalOne}{$\rho=100$}
\newcommand{\DetectorFrameChirpMassSignalOne}{$\mathcal{M}_z=1.263\,M_\odot$}
\newcommand{\MassRatioSignalOne}{$\eta = 0.2497$}

\newcommand{\SNRSignalTwo}{$\rho=100$}
\newcommand{\DetectorFrameChirpMassSignalTwo}{$\mathcal{M}_z=1.2618\,M_\odot$}
\newcommand{\MassRatioSignalTwo}{$\eta=0.2456$}

\newcommand{\PhaseVaryingTwoSignalFrequencyOverlap}{2\,Hz}
\newcommand{\ChirpMassVaryingTwoSignalFrequencyOverlap}{2\,Hz}
\newcommand{\ChirpMassVaryingTwoSignalPhaseDifference}{0}

\newcommand{\FigureTwoBlackSignalStartMinusEarliestOverlappingSignalStart}{$11.5$~hours}
\newcommand{\FigureTwoLastOverlappingSignalEndMinusBlackSignalEnd}{$8.5$~hours}
\newcommand{\FigureTwoTargetSignalDuration}{$3.5$~hours}
\newcommand{\FigureTwoTotalSignalsDuration}{$20$~hours}

\newcommand{\TargetSignalParameters}{$\mathcal{M}_z = 1.2917\,M_{\odot}$, $\eta = 0.2497$, $\chi_{1z} = \chi_{2z} = 0$}
\newcommand{\TargetSignalSNR}{$\rho = 320$}
\newcommand{\NumberOfDaysSimulated}{$35$}
\newcommand{\NumberOfOverlappingSignals}{$360$}
\newcommand{\MostSimilarChirpMass}{$1.5026\,M_{\odot}$}

\newcommand{\SmallChirpMassRatio}{$0.01\,M_{\odot}$}
\newcommand{\SmallerChirpMassRatio}{$0.001\,M_{\odot}$ }
\newcommand{\SmallChirpMassRatioFrequency}{$40\,$Hz}

\newcommand{\PercentOfSignalsWithSmallChirpMassRatio}{$2.45$\%}
\newcommand{\PercentOfSignalsWithSmallerChirpMassRatio}{$0.004$\%}
\newcommand{\PercentOfSignalsDetectableCEET}{$50$\%}

\newcommand{\PercentOfSignalsWithTwoOverlapsWithSmallChirpMassRatio}{$0.05$\%}
\newcommand{\NumberOfOverlaps}{Two}
\newcommand{\WorstCaseScenario}{2}
\newcommand{\NumberOfWorstCaseScenario}{19}

\newcommand{\NumberOfSimulatedSignals}{$1\times10^6$}
\newcommand{\ParameterUncertaintySimulatedSignals}{0.1\%}


\title{Source Confusion from Neutron Star Binaries in Ground-Based Gravitational Wave Detectors is Minimal}

\author{Aaron~D. Johnson} 
\email{aaronj@caltech.edu} 
\CIT

\author{Katerina Chatziioannou} 
\email{kchatziioannou@caltech.edu} 
\CIT 
\CITLab

\author{Will M. Farr}
\email{wfarr@flatironinstitute.org}
\CCA
\StonyBrook

\date{\today}

\begin{abstract}
Upgrades beyond the current second generation of ground-based gravitational wave detectors will allow them to observe tens of thousands neutron star and black hole binaries. 
Given the typical minute-to-hour duration of neutron star signals in the detector frequency band, a number of them will overlap in the time-frequency plane resulting in a nonzero cross-correlation. 
We examine \emph{source confusion} arising from overlapping signals whose time-frequency tracks cross. 
Adopting the median observed merger rate of $100$\,Gpc$^{-3}$yr$^{-1}$, each neutron star binary signal overlaps with an average of \PopulationAvgOverlapsFromTwoHzMedianRate{}(\PopulationAvgOverlapsFromFiveHzMedianRate{})[\PopulationAvgOverlapsFromTenHzMedianRate{}] other signals when observed from 2(5)[10]\,Hz. 
The vast majority of overlaps occur at low frequencies where the inspiral evolution is slow: \PercentOverlapsBelowFiveHz{}\% of time-frequency overlaps occur in band below 5\,Hz. 
The combined effect of overlapping signals does not satisfy the central limit theorem and source confusion cannot be treated as stationary, Gaussian noise: on average \AverageTFNumberOccupiedAtTwoHzMedianRate{}(\AverageTFNumberOccupiedAtFiveHzMedianRate{})[\AverageTFNumberOccupiedAtTenHzMedianRate{}] signals are present in a single adaptive time-frequency bin centered at 2(5)[10]\,Hz.
We quantify source confusion under a realistic neutron star binary population and find that parameter uncertainty typically increases by less than \TypicalParameterUncertaintyIncrease{} unless there are overlapping signals whose detector-frame chirp mass difference is $\lesssim$ \SmallChirpMassRatio{} and the overlap frequency is $\gtrsim$ \SmallChirpMassRatioFrequency{}.
Out of \NumberOfSimulatedSignals{} simulated signals, \PercentOfMillionSignals{} fall within this region of detector-frame chirp mass differences, but their overlap frequencies are typically lower than \SmallChirpMassRatioFrequency{}.
Source confusion for ground-based detectors, where events overlap instantaneously is significantly milder than the equivalent LISA problem, where many classes of events overlap for the lifetime of the mission.
\end{abstract}

\maketitle

\section{Introduction}

Planned improvements and upgrades of ground-based gravitational wave (GW) detectors will expand both their detection horizon and their sensitive frequency range~\cite{LIGO:2020xsf,Reitze:2019iox,Hild:2010id}. 
The expanded horizon leads to detection of neutron star (BNS) and black hole (BBH) binaries to larger distances thus increasing the detection rate by orders of magnitude. 
The increased bandwidth leads to observation times that reach hours and minutes for BNSs and BBHs respectively. 
The combined outcome of these two effects is that multiple signals will overlap in time and frequency in the data streams, leading to source confusion. 
As discussed in~\cite{Cutler:2005qq,Wang:2023ldq} and proven analytically in App.~\ref{spa-toymodel}, however, the relevant condition is not whether two signals overlap in time or frequency \emph{only}, but rather whether they overlap \emph{simultaneously} in both, i.e., if their time-frequency tracks cross. 
We therefore define \emph{overlapping signals} as those whose time-frequency tracks cross, resulting in a non-zero cross-correlation.\footnote{The cross-correlation is defined as the noise-weighted inner product between two signals, Eq.~\eqref{eq:integrand}. In App.~\ref{spa-toymodel} we analytically prove under the stationary phase approximation that the integral is nonzero if and only if two signals overlap in time and frequency simultaneously. This integral is sometimes also referred to as the ``overlap integral" leading to the confusing definition: overlapping signals are those whose overlap is nonzero.} \emph{Signal confusion} is then the effect of overlapping signals on inference. 

Overlapping signals is not a new problem for GW astronomy. 
The planned LISA mission~\cite{2017arXiv170200786A} will observe (among other sources) tens of millions of galactic white dwarf binaries, thousands of which will be individually resolvable with the rest contributing to the unresolvable Gaussian noise~\cite{1990ApJ...360...75H}.
However, the ground-based and LISA overlapping source problems are not identical.
The vast majority of LISA's white dwarf binaries have negligible frequency evolution during the mission lifetime. 
As a result, two signals that overlap in frequency at one time, will continue doing so practically indefinitely. 
BNSs as observed by ground-based detectors, on the other hand, are transient sources with strong frequency evolution. 
Two signals that overlap temporally over a long time will only overlap in both time and frequency \emph{instantaneously}. 
Moreover, the frequency evolution is faster at higher frequencies, suggesting that most overlaps occur at low frequencies. 
While one LISA white dwarf binary overlaps with another binary indefinitely, one ground-based BNS overlaps with a large number of BNSs each momentarily and preferentially at lower frequencies. The latter resembles more the case of a single massive BBH that overlaps with multiple white dwarf binaries as it sweeps through the LISA frequency band~\cite{Robson:2017ayy}.

This general picture suggests that source confusion is qualitatively different across detectors and astrophysical sources.
Through a Fisher formalism, \citet{Crowder:2004ca} showed that inference accuracy for a single white-dwarf binary in LISA deteriorates \emph{exponentially} with the number of overlapping sources.
Moving up to the decihertz range, \citet{Cutler:2005qq} showed that BNS source confusion in the Big Bang Observer instead grows as the \emph{square root} of the number of overlapping sources. 
The scaling difference is exactly due to the fact that white-dwarf binaries overlap in time-frequency over a long time~\cite{Crowder:2004ca}, while frequency evolution makes BNS overlaps momentary~\cite{Cutler:2005qq}.
Each BNS time-frequency intersection happens at a random phase; it is therefore a random walk that adds incoherently.
A separate but related question is whether overlapping signals add up to Gaussian noise.
\citet{Racine:2007gv} argued that the answer depends both on the number of sources and on the type of signal we are targeting on top of all other signals. The latter determines how far in the tails of the noise distribution we have to go for detection, i.e., to what $\sigma$ level the central limit theorem has to be satisfied. 
In the context of LISA, BBH signals are ``sufficiently different" from white-dwarf binaries that source confusion can indeed be treated as Gaussian noise~\cite{Racine:2007gv}.

Moving further up to the ground-based detector frequency range, a comparison between the astrophysical rate and observable duration reveals that multiple BNSs will be simultaneously present in the data time-series~\cite{Regimbau:2009rk,LIGOScientific:2017zlf,Pizzati:2021apa,Samajdar:2021egv,Relton:2021cax}.
Since signals, however, overlap mostly at low frequencies and ``separate" as they approach merger, current detection techniques can identify them~\cite{Regimbau:2012ir,Meacher:2015rex,Wu:2022pyg,Relton:2022whr,Miller:2023rnn} and measure their coalescence time to ${\cal{O}}(10)$\,ms~\cite{Meacher:2015rex}.
Ignoring the presence of overlapping signals can lead to parameter biases for $2$ signals that merge sufficiently close~\cite{Pizzati:2021apa,Samajdar:2021egv,Relton:2021cax,Himemoto:2021ukb,Wang:2023ldq}.
Considering more or louder signals~\cite{Reali:2022aps}, and going beyond masses and aligned-spins~\cite{Pizzati:2021apa,Samajdar:2021egv,Relton:2021cax,Himemoto:2021ukb,Wang:2023ldq} would likely increase biases.
For example, inference of more subtle effects such as spin-precession~\cite{Relton:2021cax}, tests of General Relativity~\cite{Reali:2022aps,Wu:2022pyg,Dang:2023xkj}, the NS equation of state could be more severely affected by violations of the assumption that the data are consistent with Gaussian noise~\cite{Kwok:2021zny,Payne:2022spz}.
Quantitative conclusions about source confusion are, however, complicated by the fact that the relevant picture for ground-based detectors is time-frequency overlaps of multiple signals, rather than temporal coincidences between two signals as adopted in~\cite{Pizzati:2021apa,Samajdar:2021egv,Relton:2021cax}.
To emphasize this distinction, we refer to \emph{overlapping} signals as those whose time-frequency tracks cross, and \emph{coinciding} signals as those that exist simultaneously in the datastream.

In this study, we revisit overlapping signals in ground-based detectors and quantify source confusion.
We restrict to BNSs which are expected to be the most numerous and long-lasting binary source, thus resulting in more overlapping signals.
While BBH signals may suffer from larger source confusion \textit{when they overlap}, their lower local event rate and short duration in band suggest that this is more rare than BNSs, e.g.,~\cite{Pizzati:2021apa}.
Overall, we expect that source confusion will depend on the astrophysical population properties, the astrophysical merger rate, the detector sensitivity (affecting the detected rates), and the detector low frequency performance (affecting the signal duration).
We therefore consider different networks of proposed detectors, astrophysical populations, and merger rates as described in Sec.~\ref{sec:methodology}. 
We address two questions.

\emph{1. How much time-frequency overlap is there?} 
In Sec.~\ref{sec:populationoverlaps} we simulate data with BNSs under different astrophysical rates. 
We examine time-frequency crossings and confirm the qualitative picture described above. 
Under the median local rate of 100 $\mathrm{Gpc}^{-3}\mathrm{yr}^{-1}$~\cite{KAGRA:2021duu}, each BNS's time-frequency track crosses an average of \PopulationAvgOverlapsFromTwoHzMedianRate{}
other BNSs' tracks from $2$\,Hz. 
Due to the slow frequency evolution, the majority of overlaps occurs at low frequencies: \PercentOverlapsBelowFiveHz{}\% in band below 5\,Hz and very few above 20 Hz.
Splitting data into time-frequency bins that are adapted to the signal morphology, each bin contains on average(at most over $5$ days of observation) \AverageTFNumberOccupiedAtTwoHzMedianRate{}(\MaxTFNumberOccupiedAtTwoHzMedianRate{}) signals at 2\,Hz, dropping to \AverageTFNumberOccupiedAtTenHzMedianRate{}(\MaxTFNumberOccupiedAtTenHzMedianRate{}) signals at 10\,Hz. 
The low occupation number suggests that the central limit theorem is not satisfied and BNS source confusion in ground-based detectors is not another source of Gaussian noise. 
Though not quantified in this study, we expect this conclusion to hold when further considering BBH and mixed NSBH events given their shorter duration and lower rates.

\emph{2. What is the impact of overlapping signals on parameter estimation?} 
Again with simulated data we quantify the impact of overlapping signals on parameter inference of a target BNS of interest. 
If the overlapping signals were ignored altogether, parameter inference would be subject to systematic biases~\cite{Pizzati:2021apa,Samajdar:2021egv,Relton:2021cax,Himemoto:2021ukb,Wang:2023ldq,Antonelli:2021vwg}. 
A \emph{global fit} that simultaneously analyzes all signals would mitigate such biases.
Despite its technical complications, progress in the LISA~\cite{Littenberg:2023xpl} and ground-based~\cite{Smith:2021bqc,Langendorff:2022fzq,Janquart:2022nyz,Biscoveanu:2020gds,Alvey:2023naa} contexts suggests that such solutions could be available in the timescale of third-generation ground-based detectors. 
As such, here we instead focus on the \emph{statistical} uncertainty aspect of source confusion.
Following~\citet{Crowder:2004ca}, we use the Fisher formalism as described in Sec.~\ref{sec:confusionmethodology} and compare statistical uncertainty from data with only one signal and data with multiple overlapping signals. 
In Sec.~\ref{sec:results} we show that source confusion results in a subpercent increase in parameter uncertainty per signal unless there exist overlapping signals with detector-frame chirp masses $|\Delta \mathcal{M}_z|=|\mathcal{M}_{z2} - \mathcal{M}_{z1}| \lesssim$  \SmallChirpMassRatio{}. 
Even when events with such similar masses do overlap, parameter uncertainties increase by $\gtrsim 1\%$ only if the frequency of overlap is $\gtrsim$ \SmallChirpMassRatioFrequency{}.
Out of \NumberOfSimulatedSignals{} simulated signals, \PercentOfMillionSignals{} fall within this chirp mass threshold but at frequencies lower than \SmallChirpMassRatioFrequency{}, implying that none have significant parameter uncertainty increases.
Our results qualitatively agree with those of Ref.~\cite{Himemoto:2021ukb}, generalized over BNS populations and binary parameters.

Overall, we conclude that the confusion problem in third-generation detectors, where signals usually overlap instantaneously, will be a lot more mild than the LISA case, where signals may overlap for the entirety of the mission duration.
By exploring the parameter uncertainty increase and comparing to LISA calculations~\cite{Crowder:2004ca} we quantify this comparison and comment on the efficacy of LISA data analysis strategies for third-generation detectors.
Global fit analyses that simultaneously model all data components, including instrumental noise, astrophysical/cosmological backgrounds, and transient signals are likely to be successful for third-generation detector data as well, hopefully without loss of data~\cite{Zhong:2022ylh}.
We discuss these conclusions and elaborate upon further work in Sec.~\ref{sec:conclusions}.

\section{Detector Network and Astrophysical Populations}
\label{sec:methodology}

Source confusion depends both on the properties of the detector network and on the astrophysical properties of the signals. 
In Secs.~\ref{sec:detectors} and~\ref{sec:population} we describe the networks of future detectors and astrophysical populations we consider respectively. 

\subsection{Detector networks}
\label{sec:detectors}

\begin{figure}[t]
    \centering
    \includegraphics[width=\columnwidth]{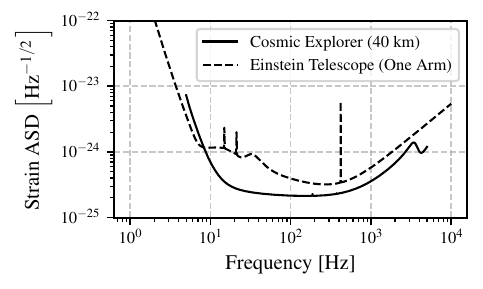}
    \caption{Projected noise amplitude spectral densities (ASDs) for the two detectors that form the different networks we consider. 
    Despite the nominal ET sensitivity going down to $1\,$Hz, we adopt a low frequency cutoff of $2$\,Hz due to the high ASD values below that frequency. Outside of these frequency ranges, the ASD is assumed to be infinite.}
    \label{fig:ASDs}
\end{figure}

We consider several detectors whose location and orientation are summarized in \autoref{tab:detectors}.
Cosmic Explorer (CE)~\cite{Reitze:2019iox} is envisioned as a 40\,km ``L'' shaped detector.
Since the location remains to be determined, we set two CE detectors at the current LIGO sites.
We adopt its projected low frequency cutoff of 5\,Hz.
While a noise curve tuned to low frequencies exists, we employ the standard noise curve for CE~\cite{evans_horizon_2021, ce_noise}, since the projected sensitivity remains the same below 10\,Hz where the majority of overlaps occur.
The Einstein Telescope (ET)~\cite{Hild:2010id} is designed with a triangular shape and 10\,km arms; we adopt the possible site of Sardinia~\cite{digiovanni_temporal_2023}.
Projected noise curves set the low-frequency sensitivity cutoff at 1\,Hz, however here we adopt a cutoff of $2\,$Hz as the noise increases rapidly below this value.
Design noise curves for all detectors are shown in Fig.~\ref{fig:ASDs}.

We combine these detectors to form different networks:
\begin{itemize}
    \item \texttt{CE}: a single CE detector in the location of LIGO's Hanford detector.
    \item \texttt{ET}: the full triangular ET detector.
    \item \texttt{CE+ET}: two CE detectors at each of LIGO's sites and an ET detector.
\end{itemize}

\subsection{Populations of neutron star binaries}
\label{sec:population}

\begin{table}[h]
    \centering
    \begin{tabular}{c|c}
    Parameter  & Prior \\\hline
    $m_1, m_2$ & $ U\left[1,2\right]M_\odot$ \\
    $\chi_{1}, \chi_{2}$ & $U\left[ -0.05, 0.05 \right]$ \\
    $\Lambda_1, \Lambda_2$  &  SFHo~\cite{Steiner:2012rk} \\
    $\cos \delta$   & $U\left[-1, 1\right]$ \\
    $\alpha$     & $U\left[0, 2\pi\right]$ \\
    $\psi$     & $U\left[0, \pi\right]$ \\
    $\cos \iota$   & $U\left[-1, 1\right]$ \\
    $t_c$      & $ U\left[0, 3024000 \right]$\,s \\
    $\phi_c$   & $ U\left[0, 2\pi \right]$ 
    \end{tabular}
    \caption{Population distributions for BNS parameters. We list the component source-frame masses $m_1,m_2$; the component spins along the orbital angular momentum $\chi_1,\chi_2$; the component dimensionless tidal deformabilities $\Lambda_1, \Lambda_2$ determined through a fixed equation of state SFHo~\cite{Steiner:2012rk} that is consistent with current observational constraints~\cite{Legred:2021hdx}; the declination $\delta$ and right ascension $\alpha$; the polarization angle $\psi$; the inclination $\iota$; the time $t_c$ and phase $\phi_c$ of coalescence. }
    \label{tab:priors}
\end{table}

We consider populations of quasicircular, spin-aligned BNS inspirals and model the GW signal with the \textsc{TaylorF2}~\cite{Buonanno:2009} waveform. 
Details about the waveform implementation and how we take the Earth's rotation into account are given in App.~\ref{app:tf2_waveform}.
Binary parameters (other than redshift) are drawn from distributions that are summarized in \autoref{tab:priors}.
Since we assume uniformly distributed masses, we adopt the corresponding local merger rates of 20, 100, and 300 $\mathrm{Gpc}^{-3}\mathrm{yr}^{-1}$ representing approximately the low, median, and high inferred values~\cite{KAGRA:2021duu}. 
Select results are also presented for the extremely high rate of 1700 $\mathrm{Gpc}^{-3}\mathrm{yr}^{-1}$ for reference.

We consider different redshift distributions computed as follows. The source-frame merger rate density is
\begin{equation}
\label{eq:merger_density}
    \dot{n}(z) \propto \int_{t_d^\text{min}}^{t_d^\text{max}}\psi\left(z_f\left(z, t_d\right)\right)P\left(t_d\right)dt_d\,,
\end{equation}
given a binary formation rate $\psi(z_f)$ and a time delay distribution $P\left(t_d\right)$ between formation and merger.
The constant of proportionality in Eq.~\eqref{eq:merger_density} is determined by matching $\dot{n}(z)$ to the local merger rate.
We assume that binary formation follows the Madau-Dickinson~\cite{madau_dickinson} star formation rate $\psi(z_f) \sim \psi_\text{SFR}(z_f)$,
\begin{equation}
\label{eq:MD}
    \psi_\text{SFR}(z_f; \alpha, \beta, z_p) = \dfrac{(1 + z_f)^\alpha}{1 + \left(\frac{1+z_f}{1 + z_p}\right)^{\alpha + \beta}}\,,
\end{equation}
with $\alpha = 2.7$, $\beta = 2.9$, and $z_p = 1.9$.
The delay time between formation and merger is
$P(t_d) \propto t_d^{-1}$.
We adopt minimum and maximum time delays of $t_d^\text{min} = 20$\,Myr and a Hubble time $t_d^\text{max} = t_H = 14.45$\,Gyr, respectively.
The mapping between redshift at formation $z_f$ and merger $z$ is obtained by solving
\begin{equation}
    t_d - \left[t_L(z_f) - t_L(z)\right] = 0\,,
\end{equation}
where
\begin{equation}
    t_L(z) = \int_0^z \frac{dz'}{(1+z')E(z')}\,,
\end{equation}
is the lookback time with
\begin{equation}
    E(z) = \sqrt{\Omega_\Lambda + \Omega_M(1 + z)^3}\,,
\end{equation}
and $\Omega_M = 0.3097$ and $\Omega_\Lambda = 0.6903$~\cite{planck_collaboration_planck_2020}.
From the rate density of Eq.~\eqref{eq:merger_density}, we obtain the rate in a redshift shell by multiplying by the comoving volume element,
\begin{equation}
    R(z) = \dot{n}(z) \frac{dV}{dz}\,.
\end{equation}
In the observer frame, $R_o(z) = R(z) / (1 + z)$.
The redshift $z$ (equivalently, luminosity distance $d_L$) distribution is,
\begin{equation}
\label{eq:redshift_dist}
    P(z) = \frac{R_o(z)}{\int_0^{\infty} R_o(z') dz'}\,.
\end{equation}
Finally, the total number of BNS events is obtained by integrating over redshift,
\begin{equation}
    N_\text{BNS} = \int_0^z R_o(z) dz'\,.
\end{equation}
\autoref{tab:pops} lists the total number of events and the average time between them for different choices of the local merger rate. 
Our results are broadly consistent with equivalent calculations using similar assumptions.
Our numbers are similar to those of~\cite{MDC,LIGOScientific:2017zlf, Sachdev:2020bkk, borhanian2022listening}, but half of those obtained in~\cite{Hu:2022bji,Wu:2022pyg} for equivalent local merger rates.
All results are highly dependent on the assumed event rates, with higher rates leading to correspondingly higher numbers of overlaps with more severe parameter estimation implications, and vice versa.
Unless otherwise noted, all subsequent results incorporate a time delay.

\begin{table}[h]
    \centering
    \begin{tabular}{c|c|c|c}
      Rate [Gpc$^{-3}$yr$^{-1}$] & Delay & $N_\text{BNS}$ & $\left<\Delta t_c\right>$ [s] \\\hline
        20 & Yes & 28955 & 1090  \\
        100 & Yes & 144778 & 218\\
        300 & Yes & 434336 & 73 \\
        1700 & Yes & 2461238 & 13 \\
        20 & No & 79694 & 396 \\
        100 & No & 398470 & 79\\
       300 & No & 1195412 & 26\\
       1700 & No & 6774004 & 5 \\
    \end{tabular}
    \caption{Simulated populations used throughout this study. We vary the local merger rate between a low, median, high, and very high value inferred in~\cite{KAGRA:2021duu} and optionally include a delay between formation and merger. The last two columns give the total number of mergers in a year $N_\text{BNS}$ and the average time between successive events $\Delta t_c$.}
    \label{tab:pops}
\end{table}

\section{The prevalence of overlapping signals}
\label{sec:populationoverlaps}

\begin{figure}[t]
    \centering
    \includegraphics[width=\columnwidth]{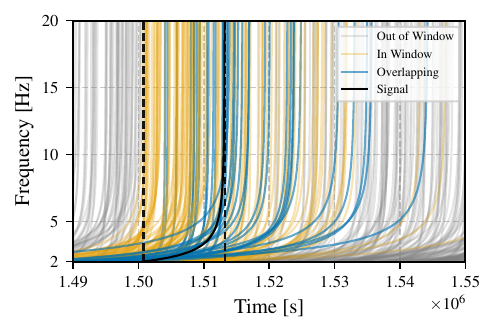}
    \caption{Approximately 2 days of simulated data for the population with the median merger rate 100 $\mathrm{Gpc}^{-3}\mathrm{yr}^{-1}$. Tracks of BNS signals through time-frequency space start with a weak frequency evolution which strengthens until the signals are nearly vertical on this scale. A target signal of interest (black) enters the band and merges within the window denoted by dashed black lines. The signal overlaps in time-frequency with all blue signals.  Orange signals exist at any point within the window, they thus coincide with the target signal only temporally. Signals never entering the window are shown in gray. Signals are far more likely to cross at lower frequencies due to the amount of time spent there. Overlapping signals can come into band far before and merge far after the target signal exists. 
    }
    \label{fig:tf_tracks}
\end{figure}

At each time, dozens of BNS signals are simultaneously present in the detector data stream~\cite{Samajdar:2021egv}. 
However, as discussed further in Sec.~\ref{sec:confusionmethodology} and App.~\ref{spa-toymodel}, source confusion is not driven by signals overlapping in time (or frequency) alone, but by signals overlapping in time and frequency simultaneously. 
In this section we study the prevalence of overlapping signals and source confusion through the time-frequency tracks of simulated signals, Eq.~\eqref{eq:timetocoa}.\footnote{A full calculation for the overlap is reserved for Sec.~\ref{sec:results}.}
Assuming each value for the local merger rate, see Table~\ref{tab:pops}, we simulate $35$ days of data with BNS signals and use the center $5$ days for the analysis.
The extra data on either side of the 5-day period reduce edge effects due to events only partially present in that period.
Figure~\ref{fig:tf_tracks} shows time-frequency tracks from a representative simulation.
Even though a large number of signals (orange) coincide with a target signal (black) temporally, a much smaller subset (blue) overlap in both time and frequency simultaneously.
The low, median, and high event rates all have a ratio of signals that overlap in time-frequency to merely coincide temporally with a target signal (number of blue signals divided by orange signals) of  $\sim$\AvgNumOverlappingOverNumInWindow{}.

\subsection{Number of time-frequency crossings per signal}
\label{sec:crossings}

Since source confusion arises from signals overlapping in time and frequency we begin by studying how often the time-frequency tracks of signals cross.
Analytical arguments inspired by \citet{Cutler:2005qq} and repeated in App.~\ref{app:Ntfcrossings} show that the rate of crossings is constant along a signal's time-frequency track, only depending on the merger rate, the BNS mass, and redshift distribution.
The number of crossings is, instead, higher at low frequencies where frequency evolution is slower.

\begin{figure}[t]
    \centering
    \includegraphics[width=\columnwidth]{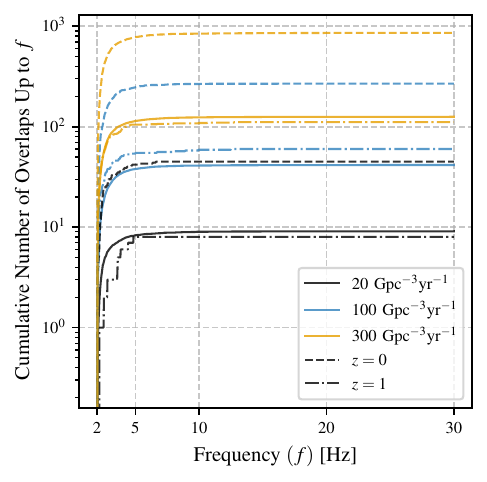}
    \caption{Cumulative number of time-frequency crossings per signal as a function of frequency. 
    We show results for a binary with the median source-frame chirp mass of \AvgChirpSource{} at $z=0$ (dashed) and at $z=1$ (dash-dotted). We also show the averaged number of crossings per signal (solid) over the entire population, a result that depends on the assumed mass and redshift distribution.
    These results are not comparable to calculations of the number of temporally coinciding signals~\cite{Relton:2021cax,Pizzati:2021apa,Samajdar:2021egv,Himemoto:2021ukb} as we consider time-frequency overlaps.
    } 
    \label{fig:overlaps}
\end{figure}

Turning to numerically simulated BNS populations, Fig.~\ref{fig:overlaps} shows cumulative numbers of crossings per signal as a function of frequency for different values of the local merger rate and BNS masses. 
Most overlaps occur below 5\,Hz. 
The number of overlaps depends sensitively on the redshift and mass of the signal under consideration, since they affect the detector-frame chirp mass and hence the signal duration. 
For example, a binary with the median source-frame chirp mass across the population of \AvgChirpSource{} undergoes \NumOverlapsAvgChirpSourceZNaught{}(\NumOverlapsAvgChirpSourceZOne{}) overlaps when observed at redshift $z=0(1)$ in the largest rate examined.
When averaged over the entire population, the number of overlaps per signal from 2 Hz is \PopulationAvgOverlapsFromTwoHzLowRate{}, \PopulationAvgOverlapsFromTwoHzMedianRate{}, \PopulationAvgOverlapsFromTwoHzHighRate{} for the low, median, and high merger rate values respectively.
The analytical calculation of App.~\ref{app:Ntfcrossings} reproduces these numerical estimates within a factor of at most $2$.

\subsection{Signals per time-frequency bin}
\label{sec:bins}

The density of signals in time-frequency space controls how large source confusion is and how difficult it is to separate the overlapping signals. 
Though a full characterization of source confusion hinges on calculating cross-correlations and likelihoods, see Sec.~\ref{sec:confusionmethodology}, we begin here by obtaining an estimate of the density of signals across time-frequency bins. 
Qualitatively, if most time-frequency bins contain at least one signal, separating signals (and characterizing the underlying stochastic noise) is more challenging.
Moreover, if the number of signals per bin is large, their combined contribution might satisfy the central limit theorem and amount to Gaussian noise, akin to LISA's white dwarf noise. 

The signal density depends on the shape of the bins, namely their time width $\delta t$ and frequency height $\delta f$ that are subject to the uncertainty (Gabor) limit~\cite{Gabor},
\begin{equation}
    \delta t \, \delta f \ge \frac{1}{2}\,.
\end{equation}
Besides this limit, the shape of the bins must be chosen to be representative of the time-frequency properties of the signals. 
For example, for a signal with constant frequency (such as a LISA white dwarf binary), increasing $\delta t$ decreases $\delta f$ and leads to improved resolution of the signal frequency. 
In other words, continued observations of the signal offers information about its frequency.
However, for a transient signal obtaining more data after the signal terminates or after it has evolved in frequency should not offer additional information.
Using a large $\delta t$ in that case would not be indicative of the frequency resolution $\delta f$ that is feasible.
The bin shapes should, therefore, be adapted to the properties of the signal and specifically the data that are relevant for the frequencies we are trying to resolve.\footnote{This tuning of the bin shapes is only important in order for the back-of-the-envelope calculation we perform here to be indicative of the full cross-correlation. In the context of a full cross-correlation or likelihood calculation, any bin shape should return the same answer. For example, a time-domain and a frequency-domain analysis should be identical if performed consistently~\cite{Isi:2021iql}.}

We adopt bins that saturate the Gabor limit $\delta t \delta f = 1/2$ and are adapted to chirps. 
At low frequencies, signals inspiral slowly hence longer (large $\delta t$, small $\delta f$) bins are optimal. 
As the signal evolves, the frequency evolution speeds up and shorter (large $\delta t$, small $\delta f$) bins are more appropriate. 
Formally, we choose bins where the ``average" signal 
enters in the lower left edge and exits in the upper right edge. This condition together with the Gabor limit uniquely define $\delta t$ and $\delta f$ at each frequency. 
An exact derivation of this condition is presented in App.~\ref{app:optimal-binning}, where we further argue that this process leads to optimal bin sizes for studying the spectral resolution and signal separation that can be achieved.

This procedure requires an ``average" signal whose detector-frame mass is used to determine the bin size. 
Besides the astrophysical population properties, the mean/median detector-frame mass depends sensitively on the detector network via its redshift reach. 
Given the uniform mass distribution, we find that the local merger rate has a minimal impact on the median detector-frame mass, varying by less than \DetectorFrameChirpMassVaryPercent{}\% between the populations of Table~\ref{tab:pops}.
Turning to the detector network, we find a median detector-frame chip mass of 
\MedianDetectorFrameChirpMassCE{}, \MedianDetectorFrameChirpMassET{}, \MedianDetectorFrameChirpMassCEET{} for, 
\texttt{CE}, \texttt{ET}, \texttt{CE+ET} respectively for signals with network signal-to-noise ratio (SNR) above 10.
Anticipating the eventual necessity of choosing a particular network, in what follows we adopt a detector-frame chirp mass of $2.8\,M_{\odot}$.
This choice impacts the following result at the level of a factor of a few.
Subsequent results continue including signals with lower SNR, but adopt this median signal for specifying the adaptive grid.

\begin{figure}[t]
    \centering
    \includegraphics[width=\columnwidth]{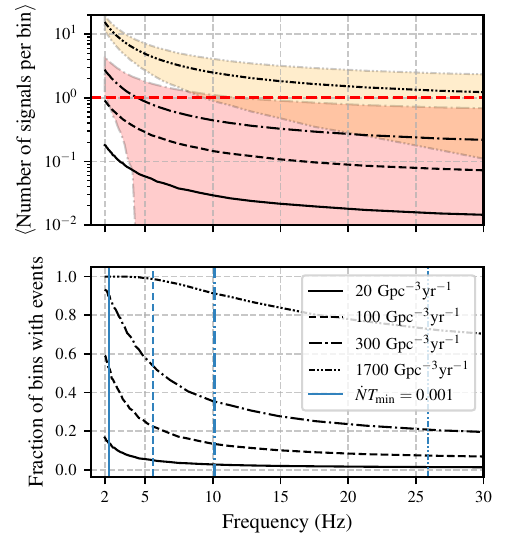}
    \caption{Average number of signals (top) and occupation fraction (bottom)
    per time-frequency bin as a function of frequency for populations
    corresponding to a low, median, high, and extremely high value of the local
    merger rate. The bin selection follows the procedure laid out in
    App.~\ref{app:optimal-binning}. The red horizontal line in the top panel
    marks the threshold where each bin on average has more than one signal.
    Shaded regions in the top panel indicate the square root of the variance of the number of signals per bin for the two highest merger rates.
    We omit shaded regions for the two lowest merger rates as they extend to zero.
    Vertical
    blue lines in the bottom panel indicate the frequency at which the
    occupation fraction has dropped below $1$ based on the analytical estimate
    from App.~\ref{app:optimal-binning}.   
    Numbers are not comparable to the estimates of~\cite{Zhong:2022ylh} who instead use a constant-size time-frequency bin. 
    }

    \label{fig:nonuniform_binning}
\end{figure}

Given the bin selection process, we simulate data, bin the time-frequency plane, and count the number of signals in each bin.
We restrict to the signal inspiral phase, as the merger and post-merger last for tens of milliseconds; they therefore do not last long enough to alter the high-frequency bin occupation where the average signal separation is on the order of a minute for the high event rate.
The bin occupation fraction is then the percentage of bins that contain at least one signal at a given frequency. 
Figure~\ref{fig:nonuniform_binning} shows the average number of signals per bin (top) and the occupation fraction (bottom) as a function of frequency for different values of the local merger rate. 
These results use all BNSs in the Universe with no cuts on SNR, thus corresponding to the worse-case scenario, and do not depend on the network. 

For the low, median, and high populations the density of signals drops sharply with frequency. 
The occupation fraction reaches 100\% at the lowest frequencies only in the extraordinarily high local rate case of 1700~$\mathrm{Gpc}^{-3}\,\mathrm{yr}^{-1}$ (at the top of the two disjoint rate ranges inferred in~\cite{KAGRA:2021duu}).
For more moderate event rates, the occupation fraction remains below $1$ and the average number of signals per bin never exceeds $\sim$10.
Above $\sim$5\,Hz the average number of signals per bin is below $1$, suggesting that source confusion is low and the stochastic noise properties (of astrophysical, cosmological, or instrumental origin) can be estimated from the empty bins.
For the low rate population, the occupation fraction remains below 20\% for all frequencies.
Even with the extraordinarily high event rate, at 2\,Hz there are on average \ExtremeEventRateAvgSignalsPerBinTwoHz{} signals per bin.\footnote{While the projected ET noise curve has a nominal low frequency cutoff of 1\,Hz, the sensitivity is reduced until around 2\,Hz where the ASD reaches $1\times 10^{-22}\,\,\mathrm{Hz}^{-1/2}$.}
These results suggest that even when considering all BNSs in the Universe (i.e., no cuts based on the signal SNR) and unrealistically high rates, the number of signals per bin does not satisfy the conditions for Gaussian noise.
Moreover, the density of signals drops sharply with frequency, suggesting that the above-threshold signals can potentially be identified and ``traced back" in frequency even in saturated bins.
We study these expectations and quantify source confusion in the remaining sections.

\section{Exploring Source Confusion within a Fisher formalism}
\label{sec:confusionmethodology}

We now turn our attention to source confusion and quantify how overlapping signals affect parameter estimation.
Since full parameter estimation via stochastic sampling of the multidimensional posterior of all signals is computationally expensive, we rely on a Fisher formalism to approximate the likelihood. 
Though such a quadratic likelihood approximation is only valid in the high SNR limit~\cite{Vallisneri:2007ev,rodriguez_inadequacies_2013} and inaqeduate for mulimodalital or complicated likelihood surfaces such as for the sky location parameters, it still corresponds to the Cramer-Rao lower bound on the variance.
Moreover, it allows us to explore source confusion under various conditions.
We adopt the methodology of \citet{Crowder:2004ca} as described below.

The data consist of stationary, zero mean, Gaussian noise $n(t)$ and a collection of $N$ signals $H(t;\vec{\Theta})$
\begin{equation}
    s(t) = H(t; \vec{\Theta}) + n(t) = \sum_{n=0}^N h_n(t; \vec{\theta}^n) + n(t)\,,
\end{equation}
where $\vec{\theta}^n$ are the parameters of signal $h_n(t; \vec{\theta}^n)$ and $\vec{\Theta}$ is a concatenation of all parameters, $\vec{\Theta} = \left(\vec{\theta}^0, \vec{\theta}^1, ..., \vec{\theta}^N\right)$.
The likelihood is then
\begin{equation}
    p(s \mid \vec{\Theta}) \propto \exp\left[{-\frac{1}{2}\left(s- H(\vec{\Theta}) \;\middle|\; s-H(\vec{\Theta}) \right)}\right]\,,
\end{equation}
where the noise-weighted inner product is
\begin{equation}
\label{eq:integrand}
    (h\mid g)=4 \operatorname{Re} \int_0^{\infty} \frac{\tilde{h}(f)^* \tilde{g}(f)}{S_n(f)} df\,,
\end{equation}
and $S_n(f)$ is the one-sided noise power spectral density (PSD).
Tildes denote Fourier transforms and ``*'' complex conjugation. 

The quadratic approximation expands each signal $h(\vec{\theta})$ around its true parameters $\vec{\theta}_t$ to first order
\begin{equation}
    h(\vec{\theta}) = h(\vec{\theta}_t) + \left.\partial_i h\right|_{\vec{\theta}_t} \, \Delta \theta^i \,,
\end{equation}
where $\Delta \theta^i \equiv \theta^i - \theta^i_t$. Equivalently for the sum of signals
\begin{equation}
    H(\vec{\Theta}) = H(\vec{\Theta}_t) + \left.\partial_i H\right|_{\vec{\Theta}_t} \, \Delta \Theta^i \,.
\end{equation}
The corresponding quadratic likelihood is
\begin{align}
    p(s \mid \vec{\Theta}) \propto \exp \left[\left(n \mid \partial_i H\right)\Delta \Theta^i-\frac{1}{2} \Gamma_{ij}\Delta \Theta^i \Delta \Theta^j\right],
    \label{eq:likelihood}
\end{align}
where $\Gamma_{ij} = \left(\partial_i H \mid \partial_j H\right)$ is the Fisher information matrix. The first term in the likelihood describes the effect of noise realization on the best-fit parameters and can be ignored if the peak of the likelihood coincides with $\Theta_t$, i.e., under a zero-noise realization. The second term describes the measurement uncertainty. The likelihood can be transformed to the posterior if augmented by some prior, $p(\vec{\Theta})$. For the wide and flat priors adopted here, the inverse of $\Gamma_{ij}$ is the covariance matrix. 

Compared to the usual Fisher matrix over a single signal, $\gamma_{ij}=\left(\partial_i h \mid \partial_j h\right)$, here we have the composite Fisher matrix that includes cross terms between signals:
\begin{equation}
    \Gamma_{ij} = \left(\frac{\partial H}{\partial \Theta^i} \;\middle|\; \frac{\partial H}{\partial \Theta^j}\right)\,,
\end{equation}
where the indices $i,j$ run over all parameters of all signals. For any signal $h_n(t;\vec{\theta}^n)$, the derivatives $\partial h_{n}(t;\vec{\theta}^n)/\partial \Theta^i$ are nonzero only if $\Theta^i \in \vec{\theta}^n$, i.e., only if we take derivatives with respect to the parameters of that signal. Moreover, inner products between different signals such as
\begin{equation}
     \left(\frac{\partial h_{n}(t;\theta^n)}{\partial \Theta^i} \;\middle|\; \frac{\partial h_{m}(t;\theta^m)}{\partial \Theta^j}\right)\,,
\end{equation}
are nonzero approximately only if the time-frequency tracks of signals $h_{n}$ and $h_{m}$ cross, as shown through the stationary phase approximation in App.~\ref{spa-toymodel}. 

The composite Fisher matrix has a block structure.
Along the diagonal, each block consists of each individual signal's Fisher matrix.
On the off-diagonal blocks, the matrix contains information about overlaps between signals.
Since the covariance is (approximately) the inverse of the Fisher matrix,
\begin{equation}
    C^{ij} = \left(\Gamma^{-1}\right)^{ij}\,,
\end{equation}
any non-zero off-diagonal pieces will affect the covariance.
Therefore, time-frequency overlaps will lead to nonzero off-diagonal terms and affect parameter uncertainties,
\begin{equation}
\label{ref:uncertainty}
    \Delta \Theta^i = (C^{ii})^{1/2}\,.
\end{equation}
These considerations motivated our study of time-frequency crossings in Sec.~\ref{sec:populationoverlaps}.

Evaluation and inversion of the composite Fisher matrix is complicated by the fact that the off-diagonal integrands are highly oscillatory and the matrix might be poorly conditioned.
Appendix~\ref{app:SPAcalculation} describes how we circumvent the first problem by analytically evaluating the off-diagonal terms under the stationary phase approximation.
Matrix condition is described by the condition number $c$, defined as the ratio of the largest to smallest eigenvalue; matrix inversion loses $\sim\log_{10}(c)$ digits of precision.
To circumvent this loss of precision and potentially numerically singular matrices, we use 100 digits of precision which allows for direct inversion of matrices with condition numbers of $\mathcal{O}(10^{20})$.
Before each inversion, we check that the Fisher matrix is positive definite and not numerically singular.
Finally, we invert with both a standard inversion method as well as a pseudo-inverse and find in practice agreement to at least 75 digits, which is far more than necessary. All calculations are performed with {\tt Mathematica}.

\section{Impact of Source Confusion on Parameter Estimation}
\label{sec:results}

In this section, we use the composite Fisher matrix from Sec.~\ref{sec:confusionmethodology} to explore how parameter uncertainty is impacted by the presence of overlapping sources.
For a ``target" signal, the parameter uncertainty ratio~\cite{Crowder:2004ca}
\begin{equation}
    \varrho = \frac{\Delta \Theta^i}{\Delta \theta^i} = \sqrt{\frac{\left(\Gamma^{-1}\right)^{ii}}{\left(\gamma^{-1}\right)^{ii}}}\,,
    \label{eq:undertaintyratio}
\end{equation}
compares the statistical uncertainty for parameter $i$ when multiple signals are present, $\Delta\Theta^{i}$, with the uncertainty when only the target signal is present, $\Delta\theta^{i}$.
The multiple-signal uncertainty $\Delta\Theta^{i}$ is obtained from the composite Fisher matrix $\Gamma$, while the single-signal uncertainty $\Delta\theta^{i}$ corresponds to only the target signal's block, $\gamma$.

\subsection{Exploring source confusion with two signals}
\label{sec:2signals}

We begin by studying the dependence of parameter uncertainty ratio on the signal parameters.
We consider two overlapping signals and all relevant data: from the time when the first signal enters the band to the time both signals have merged and from $2$\,Hz to $1024$\,Hz. The fixed upper frequency cutoff avoids biases from mass-dependent cutoffs~\cite{Mandel:2014tca}.
Under the more realistic scenario of multiple overlapping signals, considering all relevant data could result in prohibitively long datasets, c.f., Fig.~\ref{fig:tf_tracks}. 
We explore this further in Sec.~\ref{sec:BNS population}, here we focus on an exploration of the qualitative properties of source confusion.

The Fisher terms that encode source confusion are analytically computed with Eq.~\eqref{eq:overlapSPAresult} and depend on: 
\begin{enumerate}
\item the frequency at which the time-frequency tracks overlap $f_\mathrm{ov}$,\footnote{As explained in App.~\ref{app:SPAcalculation}, the stationary point of the Fisher off-diagonal terms is not exactly the same as the frequency at which the signals' time-frequency tracks cross. However, the correction is small ${\cal{O}}(10^{-2})\,$Hz, especially given that the majority of overlaps occurs at low frequencies. Though Eq.~\eqref{eq:overlapSPAresult} is computed self-consistently with the correct stationary point throughout, in our discussion we drop this distinction and refer to $f_\mathrm{ov}$.}
\item the amplitude $\mathcal{A}(f_\mathrm{ov})$ that is a combination of the signal amplitudes and phases, 
\item the phase difference between the signals $\Delta \Phi (f_\mathrm{ov})$ when they overlap,
\item and the second frequency derivative of the phase difference $d^2 \Delta \Phi/df^2(f_\mathrm{ov})$ which is related to the relative slope between the (tangents of) two signals' time-frequency tracks.
\end{enumerate}
Below we explore the impact of each of the above quantities on the source confusion.

The amplitude $\mathcal{A}(f_\mathrm{ov})$ is proportional to the individual signal amplitude and inversely proportional to the noise PSD. It therefore encodes the signals' SNRs and their SNR ratio. 
Changing the SNR ratio while keeping all detector-frame parameters constant does not affect the parameter uncertainty ratio~\cite{Crowder:2004ca}.
For two signals, this is proven by considering the inverse of a $2\times2$ block matrix in App.~\ref{app:SNRratio}: the parameter uncertainty of one signal is independent of the SNR of the other signal regardless of whether there are cross terms or not.
This conclusion appears to suggest that undetectable signals, as $\mathrm{SNR}\rightarrow0$, affect parameter uncertainties as much as loud signals.
However, this conclusion is of course incorrect~\cite{Pizzati:2021apa} and is due to the fact that the Fisher matrix is only valid in the high SNR limit, thus it does not accurately reflect the low-SNR case~\cite{Crowder:2004ca}.

To explore the effect of the overlap frequency $f_\mathrm{ov}$, the phase difference at overlap $\Delta \Phi(f_\mathrm{ov})$, and the track slope at overlap $d^2 \Delta \Phi/df^2(f_\mathrm{ov})$ we consider two signals: a ``target" signal with fixed parameters and an overlapping signals whose parameters we vary such that the quantities above change. 
The target signal has SNR \SNRSignalOne{}, no spin, detector-frame chirp mass \DetectorFrameChirpMassSignalOne{}, and symmetric mass ratio \MassRatioSignalOne.
Unless varied as described below, the overlapping signal has SNR \SNRSignalTwo{}, no spin, detector-frame chirp mass \DetectorFrameChirpMassSignalTwo{}, and symmetric mass ratio \MassRatioSignalTwo{}.
While varying $f_\mathrm{ov}$ and $\Delta \Psi(f_\mathrm{ov})$ is straightforward by time- or phase-shifting the overlapping signal, the slope $d^2 \Delta \Psi/df^2(f_\mathrm{ov})$ is more involved.
The slope of the time-frequency track depends (to leading order) on the signal detector-frame chirp mass and the frequency. While the former varies across the BNS population as we consider different masses, the latter is a property of the signal. In other words, increasing $f_\mathrm{ov}$ at fixed mass will naturally change the slope of the time-frequency track as prescribed by General Relativity and this is an effect we wish to retain as $f_\mathrm{ov}$ changes.
We therefore compute the parameter uncertainty ratio for the target signal while varying one of $f_\mathrm{ov}$, $\Delta \Phi(f_\mathrm{ov})$, or the mass at a time while keeping the other two fixed. 

\begin{figure*}[t]
    \centering
    \includegraphics[width=\textwidth]{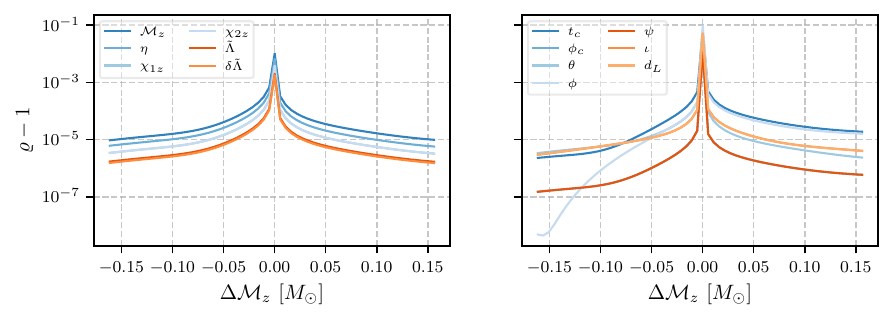}
    \caption{Uncertainty ratio $\varrho$ (minus one) for the parameters of a target signal when observed alone and when overlapping with a second signal as a function of the difference between the detector-frame chirp masses while fixing the overlap frequency to \PhaseVaryingTwoSignalFrequencyOverlap{} and the overlap phase difference $\Delta \Phi$ to \ChirpMassVaryingTwoSignalPhaseDifference{}. We consider one arm of the \texttt{ET} network and both intrinsic (left) and extrinsic parameters (right). The uncertainty increase if minimal ${\cal{O}}(10^{-5})$ unless the signals' detector-frame chirp masses differ by less than $0.01\,M_{\odot}$. 
    }
    \label{fig:overlapschirpmass}
\end{figure*}

The impact of the binary masses is explored in Fig.~\ref{fig:overlapschirpmass} which shows the parameter uncertainty ratio as a function of the difference between the detector-frame chirp mass of the target and the overlapping signal for intrinsic (left) and extrinsic (right) parameters with one arm of the \texttt{ET} network. 
The ratio is varied by changing the detector-frame chirp mass of the overlapping signal while keeping the overlap frequency (phase) constant at \ChirpMassVaryingTwoSignalFrequencyOverlap\,(\ChirpMassVaryingTwoSignalPhaseDifference). 
The increase is parameter uncertainty is negligible $\sim10^{-5}$ unless the binary masses are very similar.
Indeed, more unequal masses result in a larger relative slope between the two time-frequency tracks, thus reducing the cross-correlation.
More equal masses result in signals that remain closer in time-frequency, thus increasing the cross-correlation.
The uncertainty ratio increases to $\sim 10^{-3}$ for chirp masses that differ by less than $1\%$ and spikes to values $\geq 10\%$ across all parameters as the binary masses become even more equal.
Similar chirp masses has also been identified as a necessary condition for large confusion and biases in~\cite{Himemoto:2021ukb}.

\begin{figure*}[t]
    \centering
    \includegraphics[width=\textwidth]{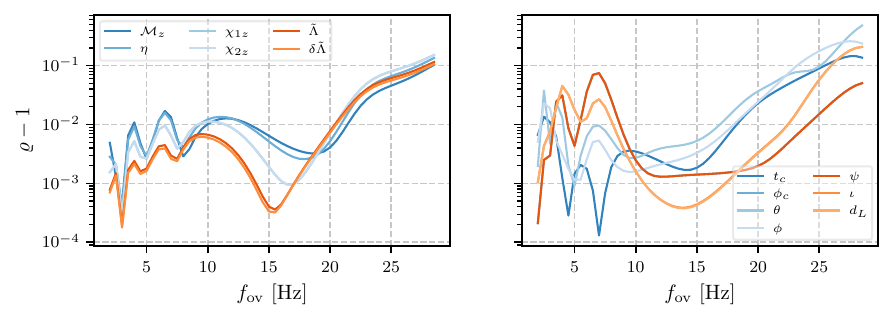}\\
    \includegraphics[width=\textwidth]{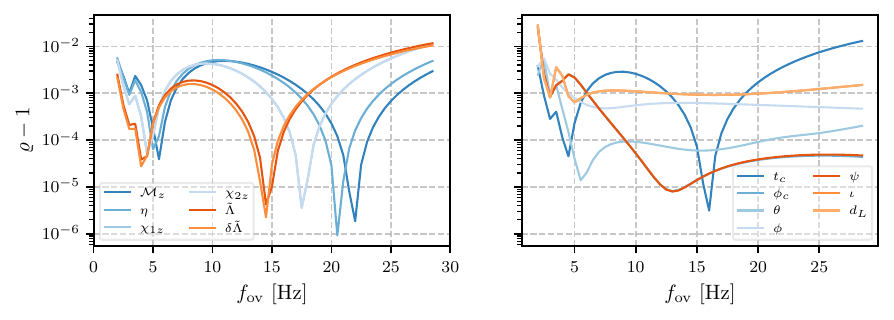}
    \caption{Uncertainty ratio $\varrho$ (minus one) for the parameters of a target signal when observed alone and when overlapping with a second signal as a function of the overlap frequency $f_\mathrm{ov}$. We consider one arm of the \texttt{ET} network (top) and a constant PSD (bottom), and both intrinsic (left) and extrinsic parameters (right). The uncertainty ratio is a sensitive function of the overlap frequency and the parameter of interest.}
    \label{fig:overlapsfrequency}
\end{figure*}
 
The impact of the overlap frequency on the parameter uncertainty ratio is explored in Fig.~\ref{fig:overlapsfrequency} for intrinsic (left) and extrinsic (right) parameters and for one arm of the \texttt{ET} network (top) and a constant PSD set to the \texttt{ET} noise curve's minimum value (bottom).
The overlap frequency is varied by shifting the overlapping signal's time of coalescence, while adjusting its phase to keep the phase difference $\Delta \Phi$ between the two signals constant at $0$ at $f_\mathrm{ov}$.
As $f_\mathrm{ov}$ increases, the difference between the two signals' track slopes, $d^2 \Delta \Phi/df^2(f_\mathrm{ov})$, decreases as the tracks steepen, leading to an increase in the uncertainty ratio for intrinsic parameters.
The overall uncertainty ratio depends sensitively on the overlap frequency and the parameter of interest.
Both intrinsic and extrinsic parameters display oscillatory behavior for frequencies $\leq20$\,Hz, with local minima at different frequency for each parameter.
At higher frequencies, the uncertainty ratio varies more smoothly, reaching $\sim 10\%$ for certain intrinsic parameters.
The PSD shape has an effect at the level of a factor of $\sim$2.

\begin{figure*}[t]
    \centering
    \includegraphics[width=\textwidth]{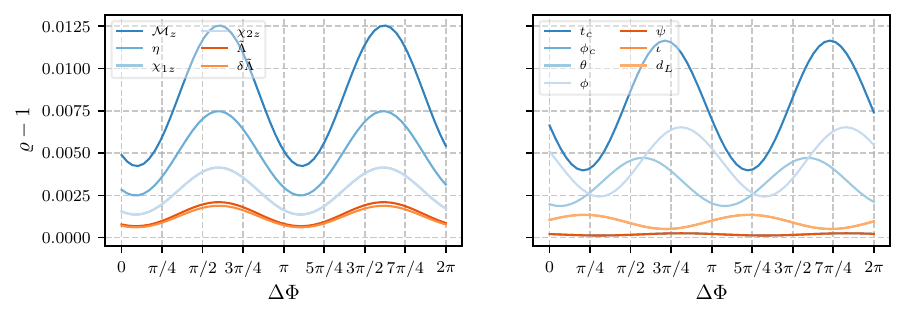}
    \caption{Uncertainty ratio $\varrho$ (minus one) for the parameters of a target signal when observed alone and when overlapping with a second signal as a function of the phase difference between two signals that overlap at \PhaseVaryingTwoSignalFrequencyOverlap{} in one arm of the \texttt{ET} network. The uncertainty ratio has a characteristic oscillatory pattern for all parameters.
    }
    \label{fig:overlapsphase}
\end{figure*}

Finally, the parameter uncertainty ratio as a function of the phase difference between the two signals when they overlap in time-frequency is shown in Fig.~\ref{fig:overlapsphase} for intrinsic (left) and extrinsic (right) parameters using a single arm of the \texttt{ET} network.
The phase difference is varied by shifting the overlapping signals' phase of coalescence, while keeping the overlap frequency constant at \PhaseVaryingTwoSignalFrequencyOverlap{}.
This procedure also keeps the difference between the two signals' track slopes $d^2 \Delta \Phi/df^2(f_\mathrm{ov})$ constant. Each parameter's uncertainty ratio shows a similar oscillatory behavior as the phase difference changes, with a period of half the GW one.
The uncertainty ratio varies with the phase difference by a factor of $\sim$2-3. 
The oscillations are in phase for all intrinsic parameters and the parameter uncertainty is maximized when the signals are approximately in phase when their tracks cross (recall the factor of $\pi/4$ in Eq.~\eqref{eq:overlapSPAresult}).

\subsection{Source confusion from BNS populations}
\label{sec:BNS population}

Figure~\ref{fig:overlapschirpmass} shows that for two overlapping signals, source confusion is negligible unless the binaries have extremely similar detector-frame chirp masses, which forces their time-frequency tracks to remain close over a range of frequencies.
In the realistic case of a BNS population, each binary overlaps with dozens or hundreds of other signals, c.f. Fig.~\ref{fig:overlaps}, and the total source confusion is their combined effect.
However, given the sharp drop of source confusion with chirp mass difference, we expect that even in the full population case, source confusion will be nonnegligible only when there exists an overlapping binary with a similar chirp mass. 

Another complication of BNS populations is that the relevant data resemble those of Fig.~\ref{fig:tf_tracks}: each signal exists within some finite time window, but it overlaps with signals that enter the detector band and merge over a wide range of times.
For example, the target signal of Fig.~\ref{fig:tf_tracks} (black) overlaps in time-frequency with signals (blue) that enter the band (merge) up to \FigureTwoBlackSignalStartMinusEarliestOverlappingSignalStart{} (\FigureTwoLastOverlappingSignalEndMinusBlackSignalEnd{}) before (after) it. 
To complicate matters further, a signal (yellow or gray) affects parameter uncertainties for the target signal (black) even if they do not overlap, if they instead both overlap with a \emph{third} signal (one of the blue signals). 
This is shown in App.~\ref{app:3signaloverlap} and suggests that in principle even a signal that enters the band weeks (or years) after the target signal affects its inferred properties.
An exact analysis would therefore have to simultaneously analyze years of data, leading to probably a prohibitive computational cost.
In practice, however, such multi-signal overlaps are subdominant to direct 2-signal overlaps unless all signals have similar chirp masses.

Even when restricting to 2-signal overlaps (the black and the blue signals in Fig.~\ref{fig:tf_tracks}), the relevant data extend over long periods of time: the target signal lasts in band for \FigureTwoTargetSignalDuration{}, while the relevant data for any overlapping signal cover a full \FigureTwoTotalSignalsDuration. 
In Sec.~\ref{sec:2signals} we presented results analyzing all the relevant data in the case of 2 overlapping signals. 
Here we adopt a more moderate setup that resembles current analysis settings~\cite{KAGRA:2021vkt}: given a target signal we wish to analyze, we consider data from the time it enters the band to the time it merges, i.e., the black-dashed window of Fig.~\ref{fig:tf_tracks}.

We considering a simulated population realization with the {\tt CE+ET} detector network.
The target signal has parameters \TargetSignalParameters{}, with $\tilde{\Lambda}$ and $\delta \tilde{\Lambda}$ given by the equation of state; its SNR is \TargetSignalSNR{}.
We consider the high inferred merger rate of 300 $\mathrm{Gpc}^{-3}\mathrm{yr}^{-1}$ and all binaries regardless of SNR. Our results are thus an upper limit on source confusion.
We simulate \NumberOfDaysSimulated{} days of data and upon injecting the target into the middle of the data, obtain \NumberOfOverlappingSignals{} signals that overlap with it. 
We then compute the composite Fisher matrix and obtain the parameter uncertainty ratio of each of the target signal parameters.

For this simulated population we find that the parameter uncertainty ratio remains very close to unity, with the largest increase at 0.2\% for the sky location parameters.
Comparing this result to Fig.~\ref{fig:overlapschirpmass} suggest that no overlapping signal has a similar chirp mass to the target. Indeed this is the case, as the most similar chirp mass in this realization is \MostSimilarChirpMass{}.
As an extension, we find that the total number of overlapping signals does not affect source confusion, the only condition being whether any of the overlapping signals has a similar detector-frame chirp mass.
Given this result, we argue that it is unlikely that BBH and NSBH signals will cause a significant increase in parameter uncertainty for BNSs, as they never have chirp masses comparable to BNS ones.
This conclusion likely applies to subthreshold BNSs as well, which typically have larger redshifts and thus are redshifted outside the typical BNS mass range.

\citet{Himemoto:2021ukb} concluded that both similar detector-frame chirp masses and coalescence time differences $|\Delta t_c|<$0.1\,s are a necessary condition for large biases. 
Our results qualitatively agree with this conclusion, as the overlapping signals we consider here will also merge close in time if they have similar chirp masses.
What is more, our results extend this conclusion to BNS populations and multiple overlapping signals.
Even when the target signal overlaps with hundreds of BNSs, it is only the signals with similar detector-frame chirp masses that result to source confusion.
The prevalence of such signals with similar masses depends sensitively on the astrophysical BNS mass distribution and is subject to Poisson uncertainty.

\begin{figure}
    \centering
    \includegraphics[width=\columnwidth]{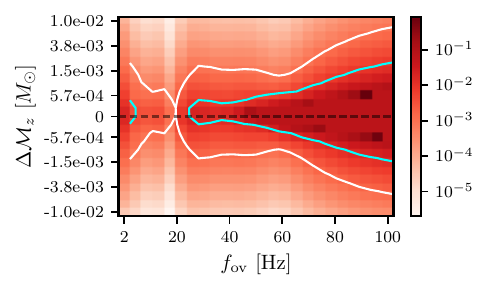}
    \caption{Geometric mean of the parameter uncertainty ratio (minus one) for all 13 BNS parameters (heat map) as a function of the overlap frequency and difference in detector-frame chirp mass between two overlapping signals. A white(cyan) contour corresponds to a 0.1\%(1\%) increase in parameter uncertainty. The horizontal dashed line indicates no difference between the two detector-frame chirp masses.}
    \label{fig:gmur}
\end{figure}

Figure~\ref{fig:gmur} explores the dependence of the parameter uncertainty ratio on the chirp mass difference in more detail. 
We show the geometric mean of the parameter uncertainty ratio (minus one) over all 13 parameters for different detector-frame chirp mass differences and overlap frequencies.
For this plot we simulate two overlapping signals at a phase difference of zero at overlap.
As expected, the parameter uncertainty ratio stays generally close to one.
Source confusion becomes more important at higher overlap frequencies, however at these frequencies we do not expect many signals overlapping to begin with, c.f., Fig.~\ref{fig:overlaps}.
Conservatively, we conclude that the parameter uncertainty ratio stays well below \TypicalParameterUncertaintyIncrease{} unless the detector-frame chirp masses differ by less than \SmallChirpMassRatio{}, see white and blue contours.
This estimate is consistent with Ref.~\cite{Himemoto:2021ukb}. 

Finally, we quantify how often signals with such similar chirp masses are expected.
Instead of simulating a fixed amount of data and thus obtaining results that depend on the assumed merger rate, we present results as a fraction of total signals.
Our results still depend on the assumed mass distribution.
We simulate a total of \NumberOfSimulatedSignals{} signals and find that \PercentOfMillionSignals{} of these fall within the detector-frame chirp mass difference of \SmallChirpMassRatio{}. However, the overlap frequencies were typically below \SmallChirpMassRatioFrequency{}, and thus too low to increase the parameter uncertainties by more than \ParameterUncertaintySimulatedSignals{}.
For reference, \NumberOfSimulatedSignals{} total signals would correspond to $3$~years of data at a merger rate of $300$\,Gpc$^{-3}$yr$^{-1}$ and 40\% of a year of data at $1700$\, Gpc$^{-3}$yr$^{-1}$ if all signals were detected.
These results are insensitive to the NS mass distribution as we obtain similar numbers for normally distributed source-frame component masses with mean $1.5\, M_{\odot}$ and standard deviation $0.2\,M_{\odot}$.
Indeed, the detector-frame chirp mass differences depend on the redshift distribution of the events more than the mass distribution.

\section{Conclusions}
\label{sec:conclusions}

In this work, we examined source confusion from overlapping BNS signals, clarifying the conditions for non-zero cross-correlation in terms of signals that overlap in both time and frequency simultaneously.
Overlapping signals appear commonly throughout the data of next-generation ground-based detectors, but with the majority of them occurring at frequencies below 5\,Hz.
Using nonuniform time-frequency bins that are adapted to BNS signals, and with common assumptions about the high-redshift population and inferred merger rates, bin occupation fraction remains below 100\% for all frequencies $\geq 2$\,Hz.
Additionally, the average number of signals per bin is $\lesssim\mathcal{O}(10)$ at 2\,Hz and falls off as frequency increases.
Even under the highest inferred merger rate of $300$\,Gpc$^{-3}$yr$^{-1}$, the combined contribution of these signals does not satisfy the conditions for Gaussian noise per time-frequency bin. A reasonable fraction of time-frequency bins contain no BNS signal and are therefore suitable for a measurement of underlying stochastic noise.
Given that BBH and NSBH signals are rarer and shorter we do not expect them to qualitatively affect this conclusion.

We then turned to quantifying source confusion, i.e., the impact of overlapping signals on statistical uncertainties for BNS parameters, extending previous literature to higher PN orders, realistic BNS populations, and more parameters.
We considered quasicircular non-precessing BNSs described with $13$ parameters under the {\tt TaylorF2} waveform model, included the Earth's rotation, extended to the lowest frequencies that next-generation detectors will be sensitive to, and considered the astrophysically expected number of overlapping signals.
Parameter uncertainty typically increases by less than \TypicalParameterUncertaintyIncrease{} unless the detector-frame chirp masses of the overlapping signals differ by less than \SmallChirpMassRatio{} and the overlap frequency is $\gtrsim$ \SmallChirpMassRatioFrequency{}.
Simulating \NumberOfSimulatedSignals{} signals, we find none of the signals with small detector-frame chirp mass differences have overlaps at high enough frequencies to have significant increases in their parameter uncertainties.
This conclusion is consistent with the analysis of~\cite{Himemoto:2021ukb}, which we have extended to a population of overlapping BNSs with the full TaylorF2 waveform from 2~Hz.
Even when there are hundreds of overlapping signals, source confusion is driven only by those that have similar masses and occur with frequency overlaps $\gtrsim$ \SmallChirpMassRatioFrequency{}.

The detector-frame chirp mass affects how closely the signals' tracks stay together in time-frequency.
As the overlap frequency increases, the signals evolve quickly through frequencies and their tracks stay closer. As an outcome, signals with more different masses can lead to source confusion, c.f., Fig~\ref{fig:gmur}, if they overlap at high frequencies.
From the population simulations, we find about 0.4\% ($\sim$4000 of $1\times 10^6$) of signals overlap at frequencies higher than 100\,Hz at any detector-frame chirp mass.
About 50\%(25\%)[7.5\%] of these signals have at least one parameter with an uncertainty increase of more than 1\%(10\%)[50\%].
Of these signals, only a fraction will be detectable. 
The extent to which this population of signals with small differences in times of coalescence can be disentangled has been considered in previous studies~\cite{Samajdar:2021egv, Janquart:2022nyz, Alvey:2023naa, Meacher:2015rex, Pizzati:2021apa, Relton:2022whr, Wang:2023ldq}.

Our results suggest that source confusion in ground-based detectors is much more mild than in LISA.
This is due to two reasons: (i) there are fewer BNSs in the ground-based detector frequency band than white-dwarf binaries in LISA's band, and (ii) at ground-based detector frequencies BNSs experience strong frequency evolution, thus most overlaps are instantaneous. 
The only exception is overlapping signals with similar detector-frame chirp masses whose time-frequency tracks overlap over an extended period of time.
Indeed, these are the only signals that can cause significant source confusion.
Future extensions to NSBH binaries or spin-precessing degrees of freedom~\cite{Apostolatos:1994} are likely to only affect these results at a quantitative level.

From a technical standpoint, such mild source confusion can likely be efficiently addressed with methods similar to LISA's global fit~\cite{Littenberg:2023xpl} without discarding data. 
Though parameter uncertainties and biases are low among BNSs populations, methods that simultaneously model all data components are still essential. 
For example, the low number of BNSs suggests that they cannot be treated as Gaussian noise~\cite{Wu:2022pyg,Reali:2022aps,Reali:2023eug}, thus estimating the actual detector noise in the non-empty bins accurately requires a concurrent treatment of astrophysical signals~\cite{Plunkett:2022zmx}. 
Another example concerns reaching the underlying cosmological stochastic background~\cite{Biscoveanu:2020gds}, without resorting to subtracting signals from the data that is bound to leave behind residuals that can mask the background~\cite{Regimbau:2016ike,Sachdev:2020bkk,Zhou:2022nmt,Zhou:2022otw, Song:2024pnk, Sharma:2020btq,Hu:2022bji,Cutler:2005qq,Robson:2017ayy,Bellie:2023jlq}. 
Finally, a full marginalization over all events is required to fully safeguard against any potential biases~\cite{Pizzati:2021apa,Samajdar:2021egv,Relton:2021cax} given the ambitious science goals of next-generation detectors: high SNR signals, spin-precession, precision constraints on General Relativity, and a measurement of the NS equation of state.
Our results suggest that source confusion is mild and can thus likely be addressed with global fit techniques developed in the timeline of next-generation detectors.

\begin{acknowledgments}

We thank Isaac Legred for providing data for the SFHo equation of state.
We thank Neil Cornish for discussions about source confusion in the context of LISA and a review of the manuscript.
We acknowledge support from the Caltech and Jet Propulsion Laboratory President’s and Director’s Fund and the Sloan Foundation.
The computations presented here were conducted in the Resnick High Performance Computing Center, a facility supported by Resnick Sustainability Institute at the California Institute of Technology.
Software used includes {\tt Mathematica}~\cite{Mathematica}, {\tt numpy}~\cite{numpy}, {\tt scipy}~\cite{scipy}, {\tt matplotlib}~\cite{matplotlib}, and {\tt astropy}~\cite{astropy2}.
A software release containing \texttt{Mathematica} notebooks for the waveform and \texttt{Python} code to simulate a cosmological population is available at \url{https://github.com/AaronDJohnson/bns\_source\_confusion}.

\end{acknowledgments}

\appendix

\section{Analytical estimate of the number of time-frequency crossings}
\label{app:Ntfcrossings}

Following the equivalent calculation in \citet{Cutler:2005qq} in the context of the Big Bang Observer, we analytically estimate the number of signals that cross the time-frequency track of a typical BNS, given an observed merger rate of $\dot{N}$. If all signals had the same detector-frame mass and ignoring higher-order modes, then the time-frequency tracks would remain parallel and not cross. So the rate of crossings $r_c$ at a given frequency depends on both the rate of signals and the relative rate of evolution of the time-frequency tracks. That is
\begin{equation}
    r_c=\frac{1}{2} \frac{dN}{df}\frac{d\Delta f}{dt}=\frac{1}{2} \rho(f) \Delta \dot{f}\,,
\end{equation}
where $\rho(f)=dN/df$ is the density of signals in frequency and $\Delta \dot{f}$ is the variation in frequency derivatives, caused by the fact that signals have different masses. The factor of $1/2$ is due to the fact that given two nearby signals there is a 50\% chance that they are diverging from each other rather than converging.

The density of signals in frequency $\rho(f)$ can be estimated from the fact that the BNS Universe is stationary, i.e., the signal rate does not change with time,
\begin{equation}
\rho(f)\frac{df}{dt} = \dot{N} = \mathrm{const.}\,,
\end{equation}
where $df/dt$ is the GW frequency evolution, Eq.~\eqref{eq:fdot}.

For the variation in frequency derivatives between signals $\Delta \dot{f}$, from Eq.~\eqref{eq:fdot} we get to leading PN order
\begin{equation}
    \frac{\Delta \dot{f}}{\dot{f}} = \frac{5}{3}\frac{\Delta {\cal{M}}_z}{{\cal{M}}_z}\,,
\end{equation}
where $\Delta {\cal{M}}_z/{\cal{M}}_z$ is the detector-frame chirp mass variation across the BNS population.

Putting everything together,
\begin{equation}
    r_c=\frac{1}{2} \rho(f) \dot{f}\frac{5}{3}\frac{\Delta {\cal{M}}_z}{{\cal{M}}_z} = \frac{5}{6} \dot{N}\left[\frac{\Delta {\cal{M}}}{{\cal{M}}} + \frac{\Delta z}{1+z}\right]\,,
    \label{eq:crossinganalytical}
\end{equation}
where ${\cal{M}}$ is the source-frame mass and $z$ is the redshift.
Assuming the fiducial uniform mass distribution of Table~\ref{tab:priors} and the redshift distribution of Eq.~\eqref{eq:redshift_dist}, $\Delta {\cal{M}}/{\cal{M}}\sim 0.14$ and $\Delta{z}/(1+z)\sim 0.41$, and hence $r_c \sim 0.46 \dot{N}$.
This rate is independent of the frequency. 
This is due to a trade off between the fact that there are more signals at low frequencies, $\rho(f)\sim f^{-11/3}$, but those signal evolve almost ``parallel" to each other and do not cross, $\delta \dot{f}\sim f^{11/3}$. 
Given a constant \emph{rate} of crossings across frequencies, the \emph{number} of crossings is higher at low frequencies where the signals spend the most time. Indeed in Sec.~\ref{sec:populationoverlaps} we numerically show that over five days of data all but one crossing occur at or below 20\,Hz for any of the event rates considered.
Using $\dot{N}$ from \autoref{tab:pops}, we find that the predicted values from Eq.~\eqref{eq:crossinganalytical} agree within a factor of less than 2, as shown in \autoref{tab:cutler_pred}.
We computed these values for an average source-frame chirp mass of \AvgChirpSource{} at redshifts of $z = 0$ and $z = 1$.

\begin{table}[h]
    \centering
    \begin{tabular}{c|c|c|c}
        Rate [Gpc$^{-3}$yr$^{-1}$] & $z$ & Analytical & Numerical  \\\hline
             20 & 0 & 28 & 49\\
             20 & 1 & 9  & 8\\
             100 & 0 & 140& 269\\
             100 & 1 & 44& 60\\
             300 & 0 & 417& 651\\
             300 & 1 & 132& 111
    \end{tabular}
    \caption{Comparison between analytical estimates based on Eq.~\eqref{eq:crossinganalytical} and the actual number of overlaps numerically computed in Sec.~\ref{sec:crossings} for a BNS with \AvgChirpSource{} at redshifts $z=0$ and $z=1$ for varying local BNS event rates.}
    \label{tab:cutler_pred}
\end{table}

\section{Optimal time-frequency binning and analytical estimate of the number of bins with a BNS}
\label{app:optimal-binning}

\begin{figure}[h]
    \centering
    \includegraphics[width=\columnwidth]{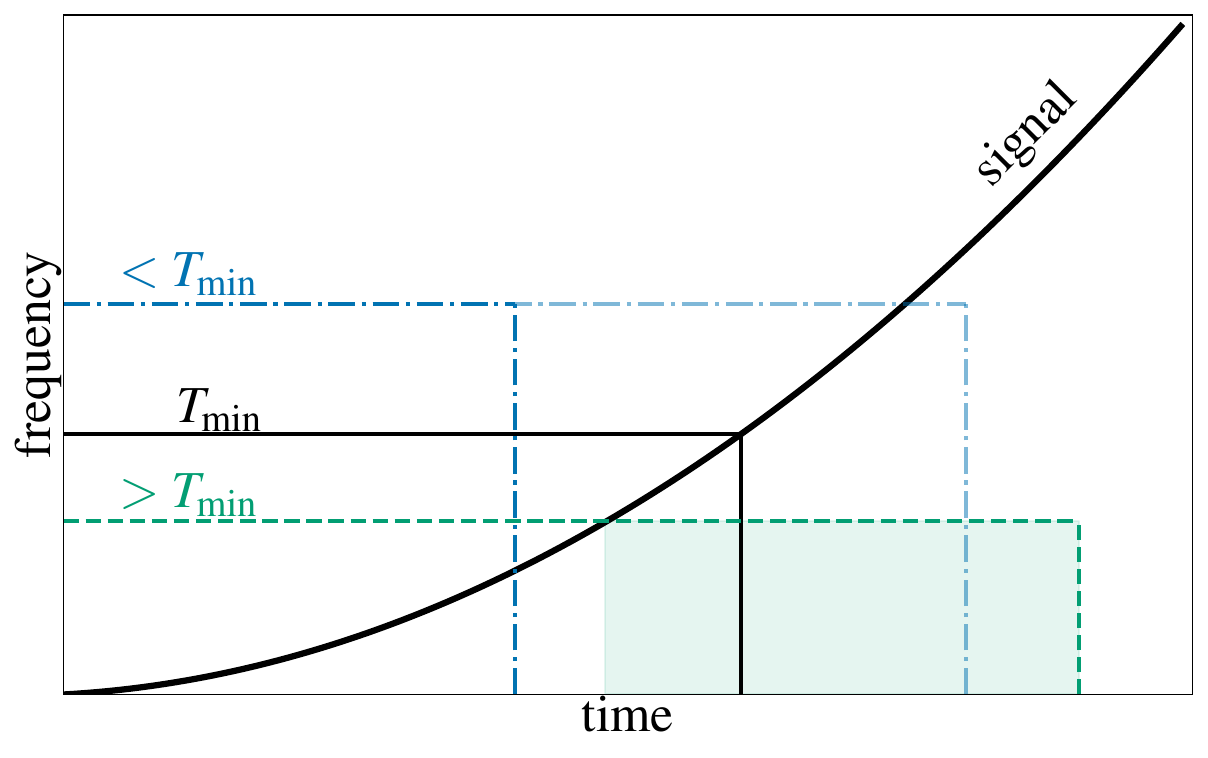}
    \caption{Visual depiction of the optimal bin selection process. 
    The time frequency track of the signal is shown in black. 
    The black box represents the optimal bin of width $T_{\rm{min}}$, whose lower-left and upper-right edges fall on the signal track. 
    This condition together with the Gabor limit define the bin dimensions. 
    A longer bin (green dashed) would contain data irrelevant to the signal (green shaded region) and thus overestimate spectral resolution.
    A shorter bin (blue dot-dashed) would leave portion of the signal to the next bin of the same frequency (pale blue dot-dashed) and thus underestimate spectral resolution. 
    A size $T_{\rm{min}}$ therefore defines the most representative bin regarding spectral resolution of a signal at a given frequency. }
    \label{fig:opt-bin}
\end{figure}

Consider BNSs entering the detector band at a rate $\dot{N}$ and merging every $T$ seconds on average. The leading-order observed frequency evolution is 
\begin{equation}
    \dot{f}=\frac{96}{5}\pi^{8/3}{\cal{M}}_z^{5/3} f^{11/3}\,,
    \label{eq:fdot}
\end{equation}
where $f$ is the observed GW frequency and ${\cal{M}}_z$ is the detector-frame chirp mass.
The minimum-time time-frequency bin that contains a signal
centered at frequency $f$ has width $T_\mathrm{min}$, where
\begin{equation}
    \delta f = \frac{1}{2 T_\mathrm{min}} \simeq \dot{f} T_\mathrm{min} \,.
    \label{eq:bin-def}
\end{equation}
In the above equation, the first equality is the Gabor condition while the second (approximate) equality enforces that the signal ``exits" the bin at its upper-right edge. 
Bins with a width larger than $T_\mathrm{min}$ would contain data after the signal has evolved in frequency and would thus overestimate how well we can resolve its frequency.
Bins with a width smaller than $T_\mathrm{min}$ would instead result in the signal also being present in the next time bin and would thus underestimate frequency resolution.
We therefore propose this time-frequency bin construction as an optimal way to assess signal spectral resolution and hence how well signals can be separated from each other.
This argument is visually presented in Fig.~\ref{fig:opt-bin}.

Equation~\ref{eq:bin-def} implies
\begin{equation}
    T_\mathrm{min} = \frac{1}{\sqrt{2 \dot{f}}} = \sqrt{\frac{5}{192}} \pi^{-4/3} \mathcal{M}_z^{-5/6} f^{-11/6}.
\end{equation}
As long as $\dot{N} T_\mathrm{min} = T_\mathrm{min} / T \ll 1$, there will be many ``empty'' time-frequency bins at frequency $f$.  This
condition is equivalent to an observed rate of signals of 
\begin{align}
    \dot{N} &\ll \frac{1}{T_\mathrm{min}} = \sqrt{\frac{192}{5}} \pi^{4/3} \mathcal{M}_z^{5/6} f^{11/6} \nonumber \\
    &= \frac{5}{\mathrm{min}} \left(\frac{{\cal{M}}_z}{1.2 \,M_\odot}\right)^{5/6} \left(\frac{f}{10 \, \mathrm{Hz}}\right)^{11/6}\,.
    \label{eq:rate-condition}
\end{align}
The vertical blue lines in the bottom panel of Fig.~\ref{fig:nonuniform_binning} show the frequency at which $\dot{N} T_\mathrm{min} = 0.001$.
The analytical estimate is a good approximation for the full numerical results, though it generally overestimates the occupation fraction of bins. We attribute this to the fact that we estimate the bin size with an ``average" signal, while the population contains a range of signal masses.

\section{Overlap integral calculation using the stationary phase approximation}
\label{spa-toymodel}

The cross-correlation, or overlap integral, between two signals is nonzero if and only if the signals cross in both time and frequency~\cite{Cutler:2005qq,Cornishpriv,Wang:2023ldq}. In this appendix we provide a proof of this through the stationary phase approximation. Without loss of generality, we restrict to a toy model with signals with linear frequency evolution.

\subsection{Signal}

We consider a sinusoidal signal with constant amplitude $A$, and a liner frequency drift. The signal phase depends on 4 parameters: the frequency $f_0$, its derivative $\dot{f}_0$, a constant time $t_0$ and a constant phase $\phi_0$ offset. Overall, $\vec{\theta}=[A,f_0,\dot{f}_0,t_0,\phi_0]$. The frequency evolution is
\begin{equation}
f(t) = f_0+\dot{f}_0(t-t_0)\,,
\end{equation}
and the signal is
\begin{equation}
h(t)=A\cos\left[2\pi\left(f_0(t-t_0) + \frac{1}{2}\dot{f}_0 (t-t_0)^2   \right)+\phi_0 \right]\,.
\end{equation}
The Fourier transform is
\begin{equation}
\tilde{h}(f) = \int h(t) e^{-2\pi i f t} dt = \int A\cos \phi(t) e^{-2\pi i f t} dt\,,
\end{equation}
where
\begin{equation}
\phi(t) = 2\pi\left(f_0(t-t_0) + \frac{1}{2}\dot{f}_0 (t-t_0)^2   \right)+\phi_0\,,
\end{equation}
is the time-domain phase. With this definition
\begin{equation}
\tilde{h}(f) = \int A\frac{e^{i\phi(t)}+e^{-i\phi(t)}}{2} e^{-2\pi i f t} dt\,.
\end{equation}
The second term has no stationary point and vanishes under the Riemann-Lebesque lemma~\cite{bender78:AMM}. The first term has a stationary point at
\begin{align}
2 \pi f_{\rm SPA} &= \dot{\phi}(t)=2\pi [f_0+\dot{f}_0(t-t_0)] \Rightarrow \nonumber  \\
f_{\rm SPA} & = f_0+\dot{f}_0(t-t_0)\,,   \\
t_{\rm SPA} & = \frac{f-f_0}{\dot{f}_0}+t_0\,.
\end{align}
With $\Phi(t)\equiv 2 \pi f t -\phi(t)$, the frequency-domain signal is
\begin{equation}
\tilde{h}(f) \approx \frac{A}{2}\int e^{-i\Phi(t)}dt \approx \frac{A}{2}\sqrt{\frac{2\pi}{|\ddot{\Phi}(t_{\rm SPA})|}}e^{-i\Phi(t_{\rm SPA})-i\pi/4}\,.
\end{equation}
Plugging in numbers we get
\begin{equation}
\tilde{h}(f) = \frac{A}{2\sqrt{\dot{f}_0}} \exp\left[-2\pi i f t_0+i\phi_0 - i\pi\frac{(f-f_0)^2}{\dot{f}_0} -i\frac{\pi}{4}\right]\,.
\label{eq:hfourier}
\end{equation}
%

\subsection{Overlapping signals}

Given two signals, the cross-correlation is
\begin{equation}
    (h_1|h_2)\equiv \Re\int \frac{h_1^*(f)h_2(f)}{S_n(f)} df\,.
\end{equation}

Substituting Eq.~\eqref{eq:hfourier}, we have
\begin{widetext}
\begin{equation}
(h_1|h_2)= \frac{A_1 A_2}{4\sqrt{\dot{f}_{0,1}\dot{f}_{0,2}}}\int \frac{ \exp\left[2\pi i f (t_{0,1}-t_{0,2})-i(\phi_{0,1}-\phi_{0,2}) + i\pi\frac{(f-f_{0,1})^2}{\dot{f}_{0,1}}- i\pi\frac{(f-f_{0,2})^2}{\dot{f}_{0,2}}\right]}{S_n(f)} df\,.
\end{equation}
\end{widetext}
The stationary point is
\begin{equation}
    f_{\rm SPA} = \frac{f_{0,2}\dot{f}_{0,1}-f_{0,1}\dot{f}_{0,2}}{\dot{f}_{0,1}-\dot{f}_{0,2}}+\frac{\dot{f}_{0,1}\dot{f}_{0,2}}{\dot{f}_{0,1}-\dot{f}_{0,2}}(t_{0,1}-t_{0,2})\,,
\end{equation}
which is reached by each signal at
\begin{equation}
    t_1(f_{\rm SPA}) = t_2(f_{\rm SPA}) = \frac{f_{0,2}-f_{0,1}}{\dot{f}_{0,1}-\dot{f}_{0,2}}+\frac{t_{0,1}\dot{f}_{0,1}-t_{0,2}\dot{f}_{0,2}}{\dot{f}_{0,1}-\dot{f}_{0,2}}\,.
\end{equation}
These times coincide, which means that the cross-correlation has a nonzero contribution only at the time when the two signals' time-frequency tracks intersect. 

\section{Waveform Implementation Details}
\label{app:tf2_waveform}

\begin{table*}[t]
    \centering
    \begin{tabular}{c|c|c|c|c|c|c|c|c}
        Detector  &  Location & Latitude $(\lambda)$ & Longitude $(\varphi)$ & Bisector Angle $(\gamma)$ & Arm Angle $(\zeta)$ & Arm Length & $f_\text{low}$ \\\hline
        ET Arm 1 &  Sardinia & $40.43^{\circ}$& $9.457^{\circ}$& $0^{\circ}$& $60^{\circ}$ & 10\,km & 2\,Hz \\
        ET Arm 2 &  Sardinia & $40.43^{\circ}$& $9.457^{\circ}$& $60^{\circ}$& $60^{\circ}$ & 10\,km & 2\,Hz \\
        ET Arm 3 &  Sardinia & $40.43^{\circ}$& $9.457^{\circ}$& $120^{\circ}$& $60^{\circ}$ & 10\,km & 2\,Hz \\
        CE H     &  Hanford    & $46.5^{\circ}$& $-119.4^{\circ}$& $171^{\circ}$& $90^{\circ}$ & 40\,km & 5\,Hz\\
        CE L     &  Livingston & $30.6^{\circ}$& $-90.8^{\circ}$& $232.7^{\circ}$& $90^{\circ}$ & 40\,km & 5\,Hz \\
    \end{tabular}
    \caption{A list of proposed detectors used in future ground-based detector networks, their locations, orientations, arm lengths, and the low-frequency cutoff used in this work.}
    \label{tab:detectors}
\end{table*}

We model BNS inspirals with the \textsc{TaylorF2} waveform approximant~\cite{Buonanno:2009} including spin~\cite{ajith_addressing_2011, mishra_ready--use_2016-1} and tidal effects~\cite{wade_systematic_2014}. Parts of our custom implementation were inspired by \texttt{GWBENCH}~\cite{borhanian_gwbench_2021}.
A BNS is described in the detector frame by parameters 
\begin{equation}
    \boldsymbol{\theta} = \{ \mathcal{M}_z, \eta, t_c, \phi_c, \chi_{1}, \chi_{2}, \Lambda_1, \Lambda_2, \theta, \phi, \psi, \iota , D_{\text{L}} \}\,,
\end{equation}
consisting of the detector-frame chirp mass, symmetric mass ratio, time of coalescence, phase at coalescence, component spins, component dimensionless tidal deformabilities, declination, right ascension, polarization angle, inclination, and luminosity distance.
Additionally, the detector location is described by
\begin{equation}
    \boldsymbol{\lambda} = \{\lambda, \varphi, \gamma, \zeta \}\,,
\end{equation}
including the latitude, longitude, angle of the detector with respect to East, and the angle between the two detector arms. Values for each detector considered here are presented in Table~\ref{tab:detectors}.
Ignoring higher-order modes, the GW amplitude in the frequency domain and under the stationary phase approximation is
\begin{equation}
    A_0(f;\boldsymbol{\theta})=\sqrt{\frac{5 \pi }{24}} \frac{\mathcal{M}_z^2}{D_\text{L}} \eta^{-7/10} u^{-7/2}\,,
    \label{eq:A0}
\end{equation}
where $u = (\pi M_z f)^{1/3}$ with the detector-frame total mass $M_z = \mathcal{M}_z\eta^{-3/5}$.
The GW phase to 3.5PN order
\begin{equation}
\begin{split}
    \Psi_{3.5\mathrm{PN}}(f;\boldsymbol{\theta},\boldsymbol{\lambda}) &= 2 \pi  f t_c - \phi_c - \frac{\pi }{4}\\
    &+ \frac{3}{128 \eta  u^5} \sum_{i=0}^7 B_i u^i  \,,
    \label{eq:Psi-fourier}
\end{split}
\end{equation}
where the $B_i$ coefficients (including log terms) exist in a \texttt{Mathematica} notebook included in our data release.

Since BNS signals can last from hours to days when observed from frequencies below $10$\,Hz, the Earth's rotation must be taken into consideration.
Our implementation follows \texttt{GWFAST}~\cite{iacovelli_forecasting_2022-1, iacovelli_gwfast_2022}.
Overall, the rotation of the Earth means that the pattern functions now depend on time~\cite{jaranowski_data_1998}.
The time to coalescence is
\begin{equation}
    \label{eq:timetocoa}
    t(f; \boldsymbol{\theta}, \boldsymbol{\lambda}) = t_c - t_*(f; \boldsymbol{\theta}) + \Delta t_\text{L}(f; \boldsymbol{\theta}, \boldsymbol{\lambda})\,,
\end{equation}
where $t_*(f; \boldsymbol{\theta})$ is Eq.~3.8b of \cite{Buonanno:2009}, and $\Delta t_\text{L}(f; \boldsymbol{\theta},\boldsymbol{\lambda})$ is the light travel time from the center of the Earth to the detector~\cite{iacovelli_gwfast_2022},
\begin{widetext}
\begin{equation}
\begin{aligned}
    \Delta t_\text{L}(f; \boldsymbol{\theta},\boldsymbol{\lambda}) = -\frac{R_\oplus}{c}&\left[\sin (\theta ) \cos (\lambda ) \cos (\phi ) \cos (2 \pi  f_\oplus (t_c - t_*(f;\boldsymbol{\theta}) + \varphi) \right. \\ &+ \sin (\theta ) \cos (\lambda ) \sin (\phi ) \sin (2 \pi  f_\oplus (t_c - t_*(f;\boldsymbol{\theta})+\varphi )\left. + \cos (\theta ) \sin (\lambda ) \right]\,,
\end{aligned}
\end{equation}
\end{widetext}
where $f_\oplus = \mathrm{day}^{-1}$.
The Earth's rotation causes amplitude and phase modulations through the time dependence of the pattern functions and Doppler shift due to the rotation of the Earth.
The frequency-domain strain is then
\begin{equation}
    \tilde{h}(f; \boldsymbol{\theta}, \boldsymbol{\lambda}) = A(f;\boldsymbol{\theta},\boldsymbol{\lambda}) e^{i \Phi(f;\boldsymbol{\theta},\boldsymbol{\lambda})}S(f) \,,
\end{equation}
where,
\begin{equation}
    A(f;\boldsymbol{\theta},\boldsymbol{\lambda}) = \sqrt{A_+^2(f;\boldsymbol{\theta},\boldsymbol{\lambda}) + A_\times^2(f;\boldsymbol{\theta},\boldsymbol{\lambda})} \,,
\end{equation}
and
\begin{align}
    A_+(f;\boldsymbol{\theta},\boldsymbol{\lambda}) &= \frac{1}{2} A_{0}(f;\boldsymbol{\theta}) F_+(f;\boldsymbol{\theta},\boldsymbol{\lambda}) \left(1 + \cos^2\iota\right)\,,\\
    A_\times(f;\boldsymbol{\theta},\boldsymbol{\lambda}) &= A_{0}(f;\boldsymbol{\theta}) F_\times(f;\boldsymbol{\theta},\boldsymbol{\lambda}) \cos \iota\,,
\end{align}
where the beam pattern functions are given in the \texttt{Mathematica} notebook.
$S(f)$ is a sigmoid function,
\begin{equation}
    S(f) = \frac{1}{1 + \exp\left(f - f_\mathrm{isco}\right)} \,,
\end{equation}
intended to create a gradual fall-off of the signal to reduce the bias produced by a hard parameter dependent frequency cutoff~\cite{rodriguez_inadequacies_2013, mandel_parameter_2014}.
The overall phase including the Earth's rotation is then
\begin{equation}
\begin{split}
    \Phi(f;\boldsymbol{\theta},\boldsymbol{\lambda}) = \Psi_{3.5\mathrm{PN}}(f;\boldsymbol{\theta},\boldsymbol{\lambda}) &+ \phi_\text{L}(f;\boldsymbol{\theta},\boldsymbol{\lambda})\\
    &+ \phi_\text{P}(f;\boldsymbol{\theta},\boldsymbol{\lambda})\,,
\end{split}
\end{equation}
where $\phi_\text{L}(f;\boldsymbol{\theta},\boldsymbol{\lambda}) = 2 \pi f \Delta t_\text{L}(f;\boldsymbol{\theta},\boldsymbol{\lambda})$ and
\begin{equation}
    \phi_\text{P}(f;\boldsymbol{\theta},\boldsymbol{\lambda}) = -\arctan\left(\frac{A_+(f;\boldsymbol{\theta},\boldsymbol{\lambda}) F_+(f;\boldsymbol{\theta},\boldsymbol{\lambda})}{A_\times(f;\boldsymbol{\theta},\boldsymbol{\lambda}) F_\times(f;\boldsymbol{\theta},\boldsymbol{\lambda})}\right)\,.
\end{equation}

\section{Fisher cross-terms in the SPA limit}
\label{app:SPAcalculation}

Non-zero block cross-terms in the composite Fisher matrix involve two signals.
In the frequency domain, cross terms correspond to the signals' phases interacting, creating a highly oscillatory integrand which cannot be easily numerically integrated.
Block diagonal Fisher terms corresponding to the same signal do not have this issue, as the phase terms are the same and conjugation results in secular integrands.
In what follows, we ignore the change in the signal at $f_\mathrm{isco}$; however, this should not effect our results since all time-frequency overlaps occur at frequencies far below $f_\mathrm{isco}$.
The cross-terms between two signals $\tilde{h}_1=A_1e^{i\Phi_1}$ and $\tilde{h}_2=A_2e^{i\Phi_2}$ are
\begin{equation}
    I = \left(\tilde{h}^{\prime}_{1} \mid \tilde{h}^{\prime}_{2}\right) = 4 \mathrm{Re}\left[\int_{-\infty}^\infty \frac{\tilde{h}^{\prime}_{1}\tilde{h}^{\prime*}_{2}}{S_n(f)} df \right]\,,
\end{equation}
where a prime denotes a derivative with respect to a single parameter.
The integrand is
\begin{equation}
    \frac{\tilde{h}^{\prime}_{1}\tilde{h}^{\prime*}_{2}}{S_n(f)}  = \frac{|\mathcal{A}|e^{i\Delta\Psi}}{S_n(f)}\,,
\end{equation}
where
\begin{equation}
    \mathcal{A} = A_1'A_2'+\Phi_1'\Phi_2'A_1A_2 + i\left(A_1A_2'\Phi_1'-A_1'A_2\Phi_2'\right)\,,
\end{equation}
\begin{equation}
    \theta = \tan^{-1}\left(\frac{A_1A_2'\Phi_1'-A_1'A_2\Phi_2'}{A_1'A_2'+\Phi_1'\Phi_2'A_1A_2}\right)\,,
\end{equation}
and
\begin{equation}
    \Delta\Psi = \Phi_1 - \Phi_2 + \theta=\Delta\Phi+\theta \,.
\end{equation}
We evaluate these derivatives symbolically.
The integrand is highly oscillatory, but it has a stationary point
\begin{equation}
    \frac{d\Delta\Psi}{df}\Big|_{f_\text{Fspa}} = 0\,.
\end{equation}
We find the stationary point $f_\text{Fspa}$ numerically.
For monotonically increasing frequencies, there is at most a single stationary point.
The integral is then
\begin{align}
    I \approx &4 \,\mathrm{Re}\left[\frac{|{\mathcal{A}}(f_\text{Fspa})|}{S_n(f)}\frac{\sqrt{2\pi}}
    {\sqrt{|d^2\Delta\Psi/df^2(f_\text{Fspa})|}}\right.\nonumber\\
    &\times
    \left.\exp\left\{i\Delta\Psi(f_\text{Fspa}) + \frac{\pi}{4}i\sign{\left[d^2\Delta\Psi/df^2(f_\text{Fspa})\right]} \right\}\right]\,.
    \label{eq:overlapSPAresult}
\end{align}
The Fisher stationary point $f_\text{Fspa}$ is not the same as the frequency at which the time-frequency tracks cross $f_\text{ov}$
\begin{equation}
    \frac{d\Delta\Phi}{df}\Big|_{f_\text{ov}}=\frac{d(\Phi_1-\Phi_2)}{df}\Big|_{f_\text{ov}} = 0\,.
\end{equation}
In practice, however, their difference is ${\cal{O}}(10^{-2})\,$Hz. Though Eq.~\eqref{eq:overlapSPAresult} is evaluated self-consistently throughout with $f_\text{Fspa}$, in the main text we largely drop the distinction between the two frequencies. 

\section{Independence of the parameter uncertainty ratio on the SNR ratio for two signals}
\label{app:SNRratio}

The composite block Fisher matrix for two signals with a set of parameters $\theta = (\mathcal{M}_z, \eta, \cdots, d_{L})$ is
\begin{equation}
    \Gamma = \begin{pmatrix}
     A& B\\
     B^{T}& D
    \end{pmatrix}\,,
\end{equation}
where $A$ is the typical Fisher matrix for the first signal only, $D$ is the typical Fisher matrix for the second signal only, and $B$ contains the overlap terms for both signals.
The inverse of this matrix is~\cite{lu_block_inverses_2002}
\begin{equation}
    \Gamma^{-1} = \begin{pmatrix}
    M^{-1} & -M^{-1} B D^{-1} \\
    -D^{-1} B^{T} M^{-1} & D^{-1}+D^{-1} B^{T} M^{-1} B D^{-1}
    \end{pmatrix}\,,
\end{equation}
where $M = A-B D^{-1} B^{T}$ is the Schur complement of the D block.
With this inverse, the parameter uncertainty ratio for the first signal is
\begin{equation}
    \varrho = \sqrt{\frac{(M^{-1})^{ii}}{(A^{-1})^{ii}}} = \sqrt{\frac{(\left(A - BD^{-1}B^{T}\right)^{-1})^{ii}}{(A^{-1})^{ii}}}\,.
\end{equation}
If we adjust the SNR of the second signal, the Fisher entries are modified:
\begin{align*}
    B' &= \alpha B \begin{pmatrix}
     1&  &  &  \\
     & 1 &  &  \\
     &  & \ddots &  \\
     &  &  & \alpha \\
    \end{pmatrix}\\
    D' &= \begin{pmatrix}
     1&  &  &  \\
     & 1 &  &  \\
     &  & \ddots &  \\
     &  &  & \alpha \\
    \end{pmatrix} \alpha^2 D \begin{pmatrix}
     1&  &  &  \\
     & 1 &  &  \\
     &  & \ddots &  \\
     &  &  & \alpha \\
    \end{pmatrix}
    \,,
\end{align*}
where $D'$ and $B'$ are the adjusted entries with modified SNR, and $\alpha$ is the factor of the SNR decrease, i.e., $\alpha = 5$ for a 5x decrease in the SNR.
The new parameter uncertainty ratio is
\begin{align}
    \varrho' &= \sqrt{\frac{(\left(A - B'D'^{-1}B'^{T}\right)^{-1})^{ii}}{(A^{-1})^{ii}}}\\
    &=\sqrt{\frac{(\left(A - BD^{-1}B^{T}\right)^{-1})^{ii}}{(A^{-1})^{ii}}} = \varrho\,,
\end{align}
it thus does not depend on $\alpha$ and the SNR ration between the two signals.

\section{Overlap between three signals}
\label{app:3signaloverlap}

Correlations propagate between non-overlapping signals if they both overlap with a third one.
Appendix~\ref{spa-toymodel} showed that the overlap between two signals has support only when the signals overlap in time and frequency. 
Consider the case where signal $1$ overlaps with signals $2$ and $3$ that do not overlap with each other.
In this case, the composite Fisher matrix is
\begin{equation}
    F= \left(\begin{matrix}
F_{11} & F_{12} & F_{13}\\
F_{12} & F_{22} & 0 \\
F_{13} & 0 & F_{33}
\end{matrix}\right) \,,
\end{equation}
where each element is a square matrix with dimensionality equal to the number of signal parameters, subscripts denote the signal,  and $F_{ij}$, $i \neq j$ represent cross terms. Since signals 2 and 3 do not overlap, $F_{23}=0$. The 33 element of the inverse is 
\begin{equation}
 F^{-1}_{33}=\frac{F_{12}^2-F_{11} F_{22}}{F_{13}^2 F_{22}+F_{12}^2 F_{33}-F_{11} F_{22} F_{33}}\,,
\end{equation}
which depends on signal 2, even though signal 3 does not overlap with it. As expected, if signals 1 and 2 did not overlap ($F_{12}=0$), the signal 2 dependence drops out of signal 3.

\bibliography{apssamp}

\end{document}